\documentclass[12pt]{article}
\usepackage[english]{babel}
\usepackage{epsfig}
\usepackage{aeguill}
\usepackage{amssymb,amsmath,xcolor}
\usepackage{amsthm}
\usepackage{graphicx}
\usepackage{natbib}
\usepackage{enumitem}
\usepackage[hidelinks]{hyperref}
\usepackage{bbm}
\usepackage{comment}
\usepackage{booktabs}

\usepackage{array}
\usepackage{makecell}
\usepackage{threeparttable}

\DeclareMathOperator*{\IF}{IF}

\newtheorem{proposition}{Proposition}
\newtheorem{example}{Example}

\newtheorem{assumption}{Assumption}

\newtheorem{lemma}{Lemma}

\begin{document}
\title{Resistant Inference in Instrumental Variable Models\footnote{We are grateful to Andreas Alfons, Isaiah Andrews, Chen Zhou, participants at the Nuffield Econometrics Lunch Seminar, European Summer Meeting of the Econometric Society (2023), International Association of Applied Econometrics conference (2023), European Winter Meeting of the Econometric Society (2022) and the Netherlands Econometric Study Group (2022) for helpful comments and suggestions.}}

\author{\normalsize Jens Klooster\footnote{Corresponding author. Department of Econometrics, Econometric Institute, Erasmus University Rotterdam, The Netherlands. E-mail: \url{Klooster@ese.eur.nl}}, Mikhail Zhelonkin\footnote{Department of Econometrics, Econometric Institute, Erasmus University Rotterdam, The Netherlands. E-mail: \url{Zhelonkin@ese.eur.nl}}}

\date{March 25, 2024}

\maketitle
\setlength{\baselineskip}{1.8\baselineskip}

\begin{abstract}
The classical tests in the instrumental variable model can behave arbitrarily if the data is contaminated. For instance, one outlying observation can be enough to change the outcome of a test. We develop a framework to construct testing procedures that are robust to weak instruments, outliers and heavy-tailed errors in the instrumental variable model. The framework is constructed upon M-estimators. By deriving the influence functions of the classical weak instrument robust tests, such as the Anderson-Rubin test, K-test and the conditional likelihood ratio (CLR) test, we prove their unbounded sensitivity to infinitesimal contamination. Therefore, we construct contamination resistant/robust alternatives.
In particular, we show how to construct a robust CLR statistic based on Mallows type M-estimators and show that its asymptotic distribution is the same as that of the (classical) CLR statistic. The theoretical results are corroborated by a simulation study. Finally, we revisit three empirical studies affected by outliers and demonstrate how the new robust tests can be used in practice.
\end{abstract}

\noindent {\it Keywords: } Influence function; Local misspecification; Robust inference; Outlier; Robust test; Weak instrument.

\newpage
\section{Introduction}\label{sec-introduction}
The linear instrumental variable (IV) model is recognized as an important tool that can be used to draw causal inferences in nonexperimental data \citep{AngristImbensRubin1996}. 
Its applicability spans over different fields and allows, for example, to infer the causal effects of education on labor market earnings \citep{angrist1991does}, chemotherapy for advanced lung cancer in the elderly \citep{earle2001effectiveness} and foreign media on authoritarian regimes \citep{kern2009opium}. In each of these examples, an endogeneity problem occurs where the explanatory variable is correlated with the error term, causing the Least Squares (LS) estimator to be biased. The researchers introduce instrumental variables to resolve this problem. The instrumental variables should be correlated with the (endogenous) explanatory variable and uncorrelated with the error term. When instrumental variables are available, then reliable estimation (and inference) is possible using IV estimators, such as the Two-Stage Least Squares (2SLS) estimator.
 
In practice, it is difficult to find instruments that satisfy both conditions. In particular, the instruments that researchers propose are oftentimes only weakly correlated with the endogenous explanatory variable \citep{andrews2019weak}. When the instruments are weak, then classical IV estimators, such as the 2SLS estimator, are biased and $t$-tests based on IV estimators can fail to control the size of the tests and their associated confidence intervals are invalid \citep{Nelson1990robust,BoundJaegerBaker1995}. 
Therefore, it is common to draw inference in the IV model using a two-step procedure. In the first step, the strength of the instruments is tested by means of an $F$-test. When the first-stage $F$ statistic is above a certain threshold (motivated by \citealt{staiger1997instrumental} a cutoff value of 10 is common), the instruments are considered to be strong. When the instruments are strong, then estimation is done with a 2SLS estimator and inference with a $t$-test. When the instruments are weak, then weak instrument robust tests are used \citep{anderson1949estimation, kleibergen2002pivotal, moreira2003conditional}.

The implementation of the two-step procedure described above by practitioners has typically been imperfect and it is advised to directly rely on weak instrument robust tests, or, smoothly adjust 2SLS $t$-test inference based on the first-stage $F$ \citep{andrews2019weak, lee2022valid, keane2023instrument}. In particular, in the just-identified model, i.e., with one endogenous variable and one instrumental variable, the \cite{anderson1949estimation} (AR) test is the uniformly most powerful among the class of unbiased tests \citep{moreira2009tests}. When there is one endogenous variable and multiple instrumental variables, then the conditional likelihood ratio (CLR) test introduced by \cite{moreira2003conditional} enjoys good power properties in the linear homoskedastic IV model \citep{andrews2006optimal}. As the CLR test is equal to the AR test in the just-identified model it is generally advised to use the CLR test. 

Recently, \cite{young2022consistency} pointed out that many empirical IV studies in economics are affected by a few outliers or by small clusters of deviating observations (see also \citealt{lal2023how} in political science). In the two-step procedure, the $F$-test is not robust against outliers \citep{ronchetti1982robust}. Hence, even one outlier is enough to inflate the first-stage $F$ leading to a false impression that the instrument is strong, while it is weak \citep{klooster2021outlier}, which eventually results in incorrect inference in the second stage. On the other hand, an outlier could also deflate the $F$-statistic, i.e., the instrument is strong, but due to an outlier it seems weak. In that case, the researcher might unnecessarily dismiss the instrument.

A possible solution could be to clean the data using outlier detection tools and then apply the (classical) procedure. However, this data-analytic strategy has been criticized in the literature, e.g., \cite{welsh2002journey}, \citet[Section~4.3]{maronna2019robust}, \citet[Section~1.3]{heritier2009robust}. If the data is cleaned, then one has to correct the inference for the data cleaning step, otherwise the classical asymptotic theory becomes invalid \citep{chen2020valid}. A typical consequence is underestimation of the (asymptotic) variance which leads to overrejection of the null hypothesis, see \cite{klooster2021outlier} for an example in the IV model. Moreover, in modern applications involving complex structural models outlier detection and its justification are themselves not trivial. Hence, a more viable strategy is to directly use (outlier) robust methods for estimation and inference. In the context of IV, several studies showed that classical IV estimators, such as the 2SLS and Limited Information Maximum Likelihood (LIML) estimators, are not robust to outliers and, consequently, introduced robust IV estimators \citep{ freue2013natural, jiao2022robust, solvsten2020robust, ZGR2012}. \cite{klooster2021outlier} developed a robust version of the AR test.
However, a general treatment of the problem of (outlier) robust inference in combination with weak instruments is missing.

In this article, we formally show that weak instrument robust tests such as the AR, K \citep{kleibergen2002pivotal} and CLR tests are not robust to outliers. For this reason, we develop a general framework that can be used to construct robust versions of the AR, K and CLR tests. In particular, we show how to construct an outlier robust CLR test that allows for reliable inference (without having to rely on a first-stage pre-test) when the data might contain some outliers. The robust CLR test provides reliable inference irrespective of the strength of the instrumental variables and only loses a little bit of power compared to the (classical) CLR test when the data does not contain any outliers. Moreover, if the errors follow a heavy-tailed distribution, then our tests are more powerful than the classical ones.

We formalize the outlier contamination using the \cite{huber1964} gross-error model $F_{t} = (1-t)F + t G$, where $t$ is a (typically small) contamination proportion. We are interested in drawing inference for the central model $F$, but we assume that it only holds approximately and the data that we observe comes from $F_{t}$. The model $G$ is assumed to be completely unknown and it is the source of the contamination. 
The goal is to draw inference that is valid at the central model $F$, but also remains stable and reliable when the data is generated according to $F_{t}$. The (classical) AR, K and CLR tests are valid when all the data is generated according to $F$, but, as we show both analytically and numerically, when the data that we observe comes from $F_{t}$, then they can be distorted. This setup is related to the literature on sensitivity analysis and local misspecification, see \cite{AndrewsGentzkowShapiro2017, AndrewsGentzkowShapiro2020}, \cite{AtheyImbens2015}, \cite{BonhommeWeidner2021}, \cite{kitamuraOtsuEvdokimov2013} and references therein. The model $F_t$ can be viewed as locally misspecified. If one assumes that the model $F$ holds exactly, then a sensitivity analysis should complement the reported results. In Section \ref{sec:case-studies}, we revisit studies by \cite{alesina2011segregation}, \cite{ananat2011wrong} and \cite{angrist1991does} and show how our tests can be used to study the sensitivity of the results to data contamination. 
Our approach allows to benefit from the parametric structure, e.g.,\ computational simplicity and interpretability, while being resistant to small but harmful deviations from the assumed model $F$. 

When the assumed model $F$ does not even hold approximately, then the use of nonparametric weak instrument robust inference \citep{andrews2007rank, andrews2008exactly} is advised. Note, however, that nonparametric procedures are typically not designed to be robust to outliers. For example, the sample mean is a nonparametric estimator of the expectation, but it is not robust as one outlier can make it arbitrarily biased (see \citealt[p.~6]{huber2009} for further discussion). Similarly, weak instrument robust quantile methods introduced by \cite{chernozhukov2008instrumental} and \cite{jun2008weak} are also not designed to be robust against outliers.

The setup of the article is as follows. In Section \ref{sec-iv}, we introduce the model and our notation. In Section \ref{sec-rob-inf}, we introduce the general (robust) framework that allows construction of weak instrument testing procedures that are robust to outliers. In particular, in Section \ref{sec:rob-clr} we show how to construct a robust version of the CLR test that can be readily used in practice. Then, in Section \ref{sec-simul-study}, we study the small sample properties of the robust CLR test and compare its performance to the classical CLR test. In Section \ref{sec:case-studies}, we revisit three empirical studies and show how the robust CLR test can be used in practice. Finally, in Section \ref{sec-conclusion} we conclude.

\section{Instrumental Variables Model} \label{sec-iv}
We assume that the data is generated according to a linear instrumental variable regression model. The model consists of a structural equation (\ref{formula-second-stage}) and a first-stage equation (\ref{formula-first-stage}):
\begin{align}
y &=  \beta x + w^{\top}\gamma_1 + u, \label{formula-second-stage} \\
x &= w^{\top} \gamma_2  + z^{\top} \pi + v, \label{formula-first-stage}
\end{align}
where $y$ and $x$ are endogenous random variables, $z$ is a $k \times 1$ random vector of instrumental variables and  $w$ is a $p \times 1$ random vector of control variables. We assume that $\gamma_1, \gamma_2 \in \mathbb{R}^p$ are parameter vectors that both, if necessary, include an intercept, $\pi \in \mathbb{R}^k$ and $\beta \in \mathbb{R}$. We assume that the errors $(u, v)$ are mean zero, with covariance matrix $\Sigma = \begin{pmatrix} \sigma_{u}^2 & \sigma_{u,v} \\ \sigma_{v, u} & \sigma_{v}^2 \end{pmatrix}$. We assume that the instruments are uncorrelated with the error terms, conditional on the control variables $w$.  

We are interested in testing the hypothesis 
\begin{align}\label{null-hypothesis-beta}
    H_0 \colon \beta = \beta_0 \text{ against } H_1 \colon \beta \neq \beta_0
\end{align} 
in the linear IV model (\ref{formula-second-stage}) - (\ref{formula-first-stage}). We do so by constructing weak instrument robust tests based on estimators in the reduced form model. The reduced form equation can be obtained by substituting (\ref{formula-first-stage}) into (\ref{formula-second-stage}): 
\begin{align}
    y &= w^{\top}\gamma + z^{\top} \delta + \epsilon, \label{for-reduced-stage2}
\end{align}
with $\gamma = (\gamma_1 +\gamma_2 \beta)$, $\delta = \pi \beta$ and $\epsilon = v\beta + u$. We refer to (\ref{formula-first-stage}) - (\ref{for-reduced-stage2}) as the reduced form model.

The tests we introduce in Section \ref{sec-rob-inf} are constructed upon estimators of the parameters $\delta$ and $\pi$ in the reduced form model (\ref{formula-first-stage}) - (\ref{for-reduced-stage2}). To simplify the presentation of our results, we assume that $\gamma_1 = \gamma_2 = 0$ so that $w$ drops out of the model. The results we present extend to the more general case where $\gamma_1 \neq 0$ and $\gamma_2 \neq 0$. In particular, when the estimators are regression equivariant (see \citealt[p.~116]{rousseeuw1987robust}), then assuming $\gamma_1 = \gamma_2 = 0$ is without loss of generality. Thus, we focus on the simplified reduced form model
\begin{align}
    y &=  z^{\top} \delta + \epsilon, \label{for-reduced-stage2-no-controls} \\
    x &=  z^{\top} \pi + v. \label{for-reduced-stage1-no-controls}
\end{align}
Define $\theta = ( \delta^{\top} , \pi^{\top} )^{\top}$, then we assume that the model (\ref{for-reduced-stage2-no-controls}) - (\ref{for-reduced-stage1-no-controls}) is governed by $F_{\theta}$. In practice, we assume that we observe i.i.d.\ data $d_i = (y_i , x_i , z_i^{\top} )^{\top}$, which are random samples from $d = (y , x , z^{\top} )^{\top} \sim F_{\theta}$. We define $Y,X \in \mathbb{R}^n$, which are vectors with entries $y_i$ and $x_i$ respectively, and $Z \in \mathbb{R}^{n \times k}$, the matrix with rows $z_i^{\top}$. Denote the partition of a vector $a \in \mathbb{R}^{2k}$ into $k$ and $k$ components by $a^{\top} = ( a_{(1)}^{\top} ,  a_{(2)}^{\top} )$ and denote the corresponding partition of $2k \times 2k$ matrices by $A_{(ij)}$, $i,j = 1,2$.

\section{Robust Inference} \label{sec-rob-inf}
In this section, we introduce a general (robust) framework to construct weak instrument robust testing procedures based on M-estimators \citep{huber1964}. We show that classical weak instrument robust tests, such as the AR, K and CLR tests can be obtained by specifying the M-estimators to be the LS estimators. To evaluate whether a statistic is robust, we use the influence function \citep{hampel1974influence, hampel1986robust}, which we discuss in the next section.

\subsection{Influence Function}\label{sec-if}
To formally determine whether a statistic is robust, we analyze its influence function. Let $T$ denote a (Fisher consistent) statistical functional, then the influence function is defined as
\begin{align}\label{for-def-if}
    \IF(d; T, F) = \lim_{t \downarrow 0} \frac{T\left\{(1-t)F + t\Delta_d\right\} - T(F)}{t},
\end{align}
where $\Delta_d$ is a point mass at $d$. The heuristic interpretation of the influence function is that it describes the effect of an infinitesimal contamination at the point $d$ on the estimate, standardized by the mass of the contamination \citep{hampel1986robust}. If the statistical functional $T$ is sufficiently regular, then a von Mises expansion \citep{mises1947asymptotic, hampel1974influence} yields
\begin{align}\label{formula-von-mises}
    T(G) \approx T(F) + \int \IF(d; T, F) d(G - F).
\end{align}
When we consider the approximation (\ref{formula-von-mises}) over a neighborhood of the model $\mathcal{F}_t = \{F_t = (1-t)F + tG \ | \ G \text{ an arbitrary distribution} \}$, we see that the influence function can be used to linearize the asymptotic bias in a neighborhood of the ideal model $F$. In particular, when we consider the maximum asymptotic bias, we obtain the following relationship
\begin{align}
    \sup_{G} || T(F_t) - T(F) || \approx t \sup_{d} || \IF(d; T, F) ||. 
\end{align}
Hence, when the influence function of a statistical functional $T$ is not bounded, then the maximum bias in a neighborhood of $F$ can be infinite, even when the contamination proportion $t$ is small. Therefore, for a statistical functional to be (locally) robust, a bounded influence function is required. 

\subsection{M-estimators}\label{sec-m-estimator}
We construct test statistics based on the M-estimator $T(F_n) = \left\{ \delta(F_n)^{\top} , \pi(F_n)^{\top} \right\}^{\top}$ of $\theta$ defined by 
\begin{equation} 
\begin{aligned} \label{for-m-est-eq}
    \frac{1}{n} \sum_{i=1}^n \Psi\left\{d_i, T(F_n)\right\} = 
        \frac{1}{n} \sum_{i=1}^n \begin{bmatrix} \psi\left\{(y_i, z_i), \delta(F_n)\right\} \\
        \psi\left\{(x_i, z_i), \pi(F_n)\right\} \end{bmatrix} = 0,
\end{aligned}
\end{equation}
where $\psi$ is a (general) score function and $F_n$ denotes the empirical distribution function. Note, letting $\psi\{(a,b),c\} = (a - b^{\top}c)b$, results in the classical LS estimating equations, which are worked out in detail in Example \ref{ex-ls-est}. 
Under general regularity conditions stated in Assumption~\ref{assumptions-main} in the Appendix, if $(y_i , x_i , z_i^{\top} )^{\top}$ are i.i.d.\ draws from the distribution $F_{\theta}$, then M-estimators are consistent and asymptotically normally distributed \citep{clarke1983uniqueness, clarke1986nonsmooth}, i.e., as $n \rightarrow \infty$, then
\begin{align*}
    \sqrt{n}\left\{T(F_n) - \theta \right\} = 
    \sqrt{n}
         \left\{\begin{matrix} 
            \delta(F_n) - \delta \\
            \pi(F_n) - \pi
         \end{matrix} \right\}
            \overset{d}{\to} \mathcal{N}\left[\begin{pmatrix} 0 \\ 0 \end{pmatrix}, \left\{\begin{matrix} \Sigma_{\delta \delta}(F_{\theta}) & \Sigma_{\delta \pi}(F_{\theta}) \\ \Sigma_{\pi \delta}(F_{\theta}) & \Sigma_{\pi \pi}(F_{\theta}) \end{matrix}\right\} \right],
\end{align*}
where
\begin{align*}
    \Sigma_{\delta \delta}(F_{\theta}) &= \int \IF\left\{d; \delta(\cdot), F_{\theta}\right\} \IF\left\{d; \delta(\cdot), F_{\theta}\right\}^{\top} dF_{\theta},\\
    \Sigma_{\pi \pi}(F_{\theta})  &= \int \IF\left\{d; \pi(\cdot), F_{\theta}\right\} \IF\left\{d; \pi(\cdot), F_{\theta}\right\}^{\top} dF_{\theta},\\
    \Sigma_{\delta \pi}(F_{\theta})  &= \int  \IF\left\{d; \delta(\cdot), F_{\theta}\right\} \IF\left\{d; \pi(\cdot), F_{\theta}\right\}^{\top} dF_{\theta},\\
    \Sigma_{\pi \delta}(F_{\theta})  &= \int \IF\left\{d; \pi(\cdot), F_{\theta}\right\} \IF\left\{d; \delta(\cdot), F_{\theta}\right\}^{\top} dF_{\theta}.
    \end{align*}
The influence functions of these estimators can be calculated using the definition (\ref{for-def-if}) (see \citealt[p.~101]{hampel1986robust}), and are given by 
\begin{align}
    \IF\left\{d_i; \delta(\cdot), F_{\theta}\right\} &= \IF\left\{d_i; T(\cdot), F_{\theta}\right\}_{(1)} = \left\{M(F_\theta)^{-1}\right\}_{(11)}  \psi\left\{(y_i,z_i), \delta\right\}, \label{for-if-m-est-delta} \\
    \IF\left\{d_i; \pi(\cdot), F_{\theta}\right\} &= \IF\left\{d_i; T(\cdot), F_{\theta}\right\}_{(2)} = \left\{M(F_\theta)^{-1}\right\}_{(22)}  \psi\left\{(x_i,z_i), \pi\right\}, \label{for-if-m-est-pi}
\end{align} 
with $M(F_\theta) = -\int \frac{\partial \Psi}{\partial \theta}(d, \theta) d F_{\theta}$.
When we analyze the influence functions (\ref{for-if-m-est-delta}) - (\ref{for-if-m-est-pi}) we see that they are only bounded when the function $\psi$ is bounded. Thus, the estimators $\delta(F_n)$ and $\pi(F_n)$ are (locally) robust when $\psi$ is bounded. 

As the LS estimators play an important role for the construction of the AR, K and CLR statistics, we explicitly show how can they be considered as a special case of the M-estimators in Example \ref{ex-ls-est}.

\begin{example}[Least Squares]\label{ex-ls-est}
    When we use the following score function:
    \begin{align*}
    \psi \colon \left(\mathbb{R} \times \mathbb{R}^k \right) \times \mathbb{R}^k \mapsto \mathbb{R}^{k} \colon \psi\{(a,b),c\} = (a - b^{\top}c)b,
    \end{align*}
     then $\delta(F_n) = (Z^{\top}Z)^{-1}Z^{\top}Y$ and $\pi(F_n)= (Z^{\top}Z)^{-1}Z^{\top}X$ are the LS estimators that solve estimating equations
    \begin{align*}
        \frac{1}{n} \sum_{i=1}^n \begin{bmatrix} \left\{y_i - z_i^{\top}\delta(F_n)\right\}z_i \\ \left\{x_i - z_i^{\top}\pi(F_n)\right\}z_i \end{bmatrix} = 0.
    \end{align*}
    Their influence functions are
    \begin{align} \label{IF:OLS}
       \IF\left\{d_i; \delta(\cdot), F_{\theta}\right\} &= \left(\int z z^{\top} dF_{\theta} \right)^{-1}\left(y_i - z_i^{\top}\delta \right)z_i, \\ 
       \IF\left\{d_i; \pi(\cdot), F_{\theta}\right\} &= \left(\int z z^{\top} dF_{\theta} \right)^{-1}\left(x_i - z_i^{\top}\pi \right)z_i. \label{IF:OLS2}
    \end{align}
    The influence functions of the LS estimators are not bounded, so that one outlying observation $d_i$ can arbitrarily bias the estimators $\delta(F_n)$ and $\pi(F_n)$.
    Using the influence function, we can now calculate the (co)variance matrices. For instance, assuming homoskedasticity, we have
    \begin{align*}
        \Sigma_{\delta \delta}(F_{\theta}) &= \int \IF\left\{d; \delta(\cdot), F_{\theta}\right\} \IF\left\{d; \delta(\cdot), F_{\theta}\right\}^{\top} dF_{\theta}\\ 
        &= \int \left(\int z z^{\top} dF_{\theta} \right)^{-1}\left(y - z^{\top}\delta \right)z z^{\top} \left(y - z^{\top}\delta \right)^{\top} \left(\int z z^{\top} dF_{\theta} \right)^{-\top}  d F_{\theta} 
        \\ 
        &= \left(\int z z^{\top} dF_{\theta} \right)^{-1} \int \left(y - z^{\top}\delta \right)^2 dF_{\theta},
    \end{align*}
    which can be estimated by
    \begin{equation*}
        {\Sigma}_{\delta \delta}(F_n) = \left(\int z z^{\top} dF_{n} \right)^{-1} \int \left\{y - z^{\top}\delta(F_n) \right\}^2 dF_{n} = \left( \frac{1}{n}Z^{\top} Z \right)^{-1}\sigma_{\epsilon}^2(F_n),
    \end{equation*}
    where $\sigma_{\epsilon}^2(F_n) = \frac{1}{n}\sum_{i=1}^n\left\{y_i - z_i^{\top}\delta(F_n) \right\}^2$.
\end{example}

\subsection{The (Robust) Conditional Likelihood Ratio Test} \label{sec-clr}
\cite{moreira2003conditional} shows that in the case of a scalar $\beta$, then the classical CLR statistic can be decomposed into the classical AR, K, and a Wald statistic. Therefore, to introduce a general (robust) version of the CLR statistic, we start by introducing general (robust) versions of the AR, K and Wald statistics. Then, we plug in the general (robust) AR, K and Wald statistics into the decomposition to introduce a general (robust) CLR statistic.

 Let $\left\{ \delta(F_n)^{\top} , \pi(F_n)^{\top} \right\}^{\top}$ denote M-estimators of the parameters $( \delta^{\top} , \pi^{\top})^{\top}$  as defined in (\ref{for-m-est-eq}). Then, given a hypothesized $\beta_0$ value, we define (see also \citealt{andrews2019weak}),
\begin{align*}
    g(F_n, \beta_0) &= \delta(F_n) - \pi(F_n)\beta_0, \\
    D(F_n, \beta_0) &= \pi(F_n) - \left\{\Sigma_{\pi \delta}(F_{\theta}) - \Sigma_{\pi \pi}(F_{\theta}) \beta_0\right\}\Omega(F_{\theta},\beta_0)^{-1}g(F_n, \beta_0),
\end{align*}
where $\Omega(F_{\theta},\beta_0) = \Sigma_{\delta \delta}(F_{\theta}) - \beta_0\left\{\Sigma_{\delta \pi}(F_{\theta}) + \Sigma_{\pi \delta}(F_{\theta})\right\} + \beta_0^2 \Sigma_{\pi \pi}(F_{\theta})$. The general (robust) AR, K and Wald statistics are defined as:
\begin{align}
    RAR(F_n, \beta_0) &= g(F_n, \beta_0)^{\top} \Omega(F_{\theta}, \beta_0)^{-1} g(F_n, \beta_0), \label{for:RAR}\\
    RK(F_n, \beta_0) &= g(F_n, \beta_0)^{\top} D(F_n, \beta_0) \left\{D(F_n, \beta_0)^{\top} \Omega(F_{\theta}, \beta_0) D(F_n, \beta_0) \right\}^{-1} D(F_n, \beta_0)^{\top} g(F_n, \beta_0), \label{for:RK}\\
    RW(F_n, \beta_0) &= D(F_n, \beta_0)^{\top} \Lambda(F_{\theta}, \beta_0)^{-1} D(F_n, \beta_0), \label{for:RW}
\end{align}
where $\Lambda(F_{\theta},\beta_0) = \Sigma_{\pi \pi}(F_{\theta}) - \left\{\Sigma_{\pi \delta}(F_{\theta}) - \Sigma_{\pi \pi}(F_{\theta}) \beta_0\right\}\Omega(F_{\theta},\beta_0)^{-1}\left\{\Sigma_{\delta \pi}(F_{\theta}) - \Sigma_{\pi \pi}(F_{\theta}) \beta_0\right\}.$ The general (robust) CLR statistic can then be written as:
\begin{equation}
\begin{aligned}\label{for-clr-stat}
    RCLR(F_n, \beta_0) =\ & \frac{1}{2}\bigg[RAR(F_n, \beta_0) - RW(F_n, \beta_0)\ + \\
      & \ \sqrt{\left\{RAR(F_n, \beta_0) - RW(F_n, \beta_0)\right\}^2 + 4RW(F_n, \beta_0) \cdot RK(F_n, \beta_0) } \bigg].
\end{aligned}
\end{equation}

In Example~\ref{ex-clr-ols}, we show how plugging in the LS estimators from Example~\ref{ex-ls-est} results in the classical AR, K, Wald and CLR statistics.

\begin{example}[Least Squares (continued)] \label{ex-clr-ols}
    When $\delta(F_n)$ and $\pi(F_n)$ are the LS estimators, and using the estimators $\Omega(F_n, \beta_0)$ and $\Lambda(F_n, \beta_0)$, then we have 
    \begin{align*}
        g(F_n, \beta_0) &= (Z^{\top} Z)^{-1}Z^{\top}(Y - \beta_0 X), \\
        D(F_n, \beta_0) &= (Z^{\top}Z)^{-1}Z^{\top}\left\{X - \frac{\sigma_{u,v}(F_n)}{\sigma_u^2(F_n)}\left(Y - X\beta_0\right)\right\}.
    \end{align*}
     Plugging these into the statistics (\ref{for:RAR}), (\ref{for:RK}) and (\ref{for:RW}) yields the classical AR, K and Wald statistics (up to a factor $n$):
    \begin{align*}
        n AR(F_n, \beta_0) &= \frac{(Y - \beta_0 X)^{\top} Z (Z^{\top}Z)^{-1} Z^{\top}(Y - \beta_0 X)}{\sigma_u^2(F_n)},\\
    nK(F_n, \beta_0) &= \frac{1}{\sigma_u^2(F_n)}(Y-X\beta_0)^{\top} P_{ZD(F_n, \beta_0)}(Y-X\beta_0),\\
    nW(F_n, \beta_0) &= \left\{X - \frac{\sigma_{u,v}(F_n)}{\sigma_u^2(F_n)}\left(Y - X\beta_0\right) \right\}^{\top}\left[\left\{\sigma_v^2(F_n) -\frac{\sigma_{u,v}(F_n)}{\sigma_u^2(F_n)}\right\}Z^{\top}Z\right]^{-1}\times \\ &  \qquad   \left\{X - \frac{\sigma_{u,v}(F_n)}{\sigma_u^2(F_n)}\left(Y - X\beta_0\right)\right\},
    \end{align*}
with $P_{ZD(F_n, \beta_0)} = ZD(F_n, \beta_0)\left\{D(F_n, \beta_0)^{\top}Z^{\top}ZD(F_n, \beta_0)\right\}^{-1}D(F_n, \beta_0)^{\top}Z^{\top}$ and $\sigma_u^2(F_n)= \sigma_{\epsilon}^2(F_n) - 2\beta_0 \sigma_{\epsilon, v}(F_n) + \beta_0^2 \sigma_v^2(F_n)$. Plugging these statistics into the decomposition (\ref{for-clr-stat}) yields the CLR statistic introduced by \cite{moreira2003conditional}.
\end{example}

{\it Remark} \ \ We are not the first to use the decomposition (\ref{for-clr-stat}) to construct a generalization of the CLR statistic. For example, \cite{kleibergen2005testing} uses the decomposition to introduce a generalized method of moments (GMM) version of the CLR statistic based on GMM versions of the AR \citep{stock2000gmm} and K statistic. \cite{magnusson2010inference} uses the decomposition to introduce a CLR test for inference in weakly identified limited dependent variable models. Their goal is to provide tests that can be used for inference in weakly identified models beyond the linear IV model. In this article, we restrict our focus to the linear IV model and leave connections to more general versions open for future research. Note, \cite{ronchetti2001robust} study the robustness properties of GMM estimators, but they do not consider weak identification.

\subsection{Robustness Properties}\label{sec-rob-issues}
To determine the robustness properties of the general (robust) CLR statistic we derive its influence function. It turns out that directly applying the influence function defined in (\ref{for-def-if}) to the statistical functional $RCLR(\cdot, \beta_0)$ is not meaningful as the influence function is always zero. This happens, because the statistical functional $RCLR(\cdot, \beta_0)$ is not Fisher consistent. A functional $T$ is Fisher consistent when $T(F_{\theta}) = \theta$ for all $\theta$, while $RCLR(F_{\theta}, \beta) = 0$. Therefore, to still obtain a meaningful result, similar as in \citet[p.~348]{hampel1986robust}, we derive the influence function of the square root of the statistic. Note, generalizations of the influence function towards functionals that are not Fisher consistent have been introduced as well \citep{rousseeuw1981influence}.

In Proposition \ref{prop-IF-RCLR}, we derive the influence function of the general (robust) CLR statistic, conditional on $D(F_n, \beta_0) = \tilde{D}$, denoted by $RCLR(\cdot, \beta_0 | \tilde{D})$.

\begin{proposition} \label{prop-IF-RCLR}
Under the null hypothesis $\beta = \beta_0$, and the regularity conditions in Assumption~\ref{assumptions-main}, the influence function of the general (robust) CLR statistic, conditional on $D(F_n, \beta_0) = \tilde{D}$, is
\begin{align*}
     \IF\left\{d_i; \sqrt{RCLR(\cdot, \beta_0 | \tilde{D})}, F_{\theta}  \right\} = \begin{cases} \IF\left\{d_i; \sqrt{RAR(\cdot, \beta_0)}, F_{\theta} \right\}, &\text{if } \tilde{D} = 0, \\ 
     \IF\left\{d_i; \sqrt{RK(\cdot, \beta_0 |  \tilde{D})}, F_{\theta}\right\}, &\text{if } \tilde{D} \neq 0.
    \end{cases}
\end{align*}
\end{proposition}
\begin{proof}
    See Appendix.
\end{proof}
\noindent
Proposition \ref{prop-IF-RCLR} shows that the influence function of the RCLR statistic, conditional on $D(F_n, \beta_0) = \tilde{D}$, depends on the influence functions of the RAR statistic and RK statistic. This corresponds to the theory, as $D(F_n, \beta_0) = 0$ suggests that true value of $\pi = 0$ as $D(F_{\theta}, \beta) = \pi$. When $D(F_n, \beta_0) = 0$, it suggests that the instruments are completely irrelevant. In this case, the RCLR statistic is the same as the RAR statistic and therefore has the same influence function as the RAR statistic. When $\tilde{D} \neq 0$, it suggests that $\pi \neq 0$. \cite{moreira2003conditional} shows that when $\pi \neq 0$, then the CLR statistic converges towards the $K$ statistic as the amount of data goes towards infinity. For this reason, we see the same behavior in the IF of the RCLR statistic as it is equal to the IF of the RK statistic when $\tilde{D} \neq 0$.

We continue with deriving the influence functions of the RAR, RW and RK statistics.

\begin{proposition}\label{prop-IF-tests}
Under the null hypothesis $\beta = \beta_0$, and the regularity conditions in Assumption \ref{assumptions-main}, the influence functions of the general (robust) AR, W and K statistics are
\begin{align}
    \IF\{d_i; \sqrt{RAR(\cdot, \beta_0)}, F_{\theta}\} &= \sqrt{\IF\{d_i; g(\cdot, \beta_0), F_{\theta}\}^{\top} \Omega(F_{\theta}, \beta_0)^{-1}\IF\{d_i; g(\cdot, \beta_0), F_{\theta}\}}, \\
    \IF\{d_i; \sqrt{RK(\cdot, \beta_0)}, F_{\theta}\} &= \sqrt{\IF\{d_i; g(\cdot, \beta_0), F_{\theta}\}^{\top}\pi \left(\pi^{\top}\Omega(F_{\theta},\beta_0)\pi\right)^{-1}\pi^{\top}\IF\{d_i; g(\cdot, \beta_0), F_{\theta}\}},\\
    \IF\{d_i; \sqrt{RW(\cdot, \beta_0)}, F_{\theta}\} &= \sqrt{\IF\{d_i; D(\cdot, \beta_0), F_{\theta}\}^{\top} \Lambda(F_{\theta},\beta_0)^{-1}\IF\{d_i; D(\cdot, \beta_0), F_{\theta}\}},
\end{align}
where 
\begin{align*}
    \IF\{d_i; g(\cdot, \beta_0), F_{\theta}\} &= \IF\{d_i; \delta(\cdot), F_{\theta}\} - \beta_0 \IF\{d_i; \pi(\cdot), F_{\theta}\} \\
    \IF\{d_i; D(\cdot, \beta_0), F_{\theta}\} &= \IF\{d_i; \pi(\cdot), F_{\theta}\} - \left\{\Sigma_{\pi \delta}(F_{\theta}) - \Sigma_{\pi \pi}(F_{\theta}) \beta_0\right\}\Omega(F_{\theta}, \beta_0)^{-1}\IF\{d_i; g(\cdot, \beta_0), F_{\theta}\}.
\end{align*}
The influence function of the general (robust) K statistic, conditional on $D(F_n, \beta_0) = \tilde{D}$, is:
\begin{align*}
    \IF\{d_i;& \sqrt{RK(\cdot, \beta_0 | \tilde{D})} , F_{\theta} \} =  \sqrt{\IF\{d_i; g(\cdot, \beta_0), F_{\theta}\}^{\top}\tilde{D} \left\{\tilde{D}^{\top}\Omega(F_{\theta},\beta_0)\tilde{D}\right\}^{-1}\tilde{D}^{\top}\IF\{d_i; g(\cdot, \beta_0), F_{\theta}\}}.
\end{align*}
\end{proposition}
\begin{proof}
    See Appendix.
\end{proof}
\noindent
Proposition \ref{prop-IF-tests} shows that the influence functions of the RAR and RK statistics depend directly on the influence function of the functional $g(\cdot, \beta_0)$. The influence function of the functional $g(\cdot, \beta_0)$ depends on the influence functions of the functionals $\delta(\cdot)$ and $\pi(\cdot)$.

\subsubsection{Robustness Issues With the Conditional Likelihood Ratio Test}
In this section, we study the robustness properties of the classical CLR statistic by analyzing its influence function.
Example \ref{ex-clr-ols} showed us how the classical CLR statistic can be viewed as a special case where LS estimators are used. Therefore, we can use Propositions \ref{prop-IF-RCLR} and \ref{prop-IF-tests} to analyze the influence function of the classical CLR statistic. As in Example \ref{ex-clr-ols}, we denote the classical AR, K, Wald and CLR statistics as $AR(F_n,\beta_0), K(F_n, \beta_0), W(F_n, \beta_0)$ and $CLR(F_n, \beta_0)$, respectively. 

The previous section showed that the influence function of the general (robust) CLR statistic ultimately depends on the influence functions of the estimators $\delta(F_n)$ and $\pi(F_n)$ it is constructed upon. The classical CLR statistic is constructed upon LS estimators. The influence functions of the LS estimators were derived in Example~\ref{ex-ls-est}, see (\ref{IF:OLS}) and (\ref{IF:OLS2}).
These influence functions are not bounded, so that one outlying observation $d_i = (y_i , x_i , z_i^{\top} )^{\top}$ can arbitrarly bias the estimators $\delta(F_n)$ and $\pi(F_n)$. As the influence functions of $\delta(\cdot)$ and $\pi(\cdot)$ are not bounded, the influence function of $g(\cdot, \beta_0)$ is not bounded, so that the influence functions $\IF\{d_i; \sqrt{AR(\cdot, \beta_0)}, F_{\theta}\}$ and $\IF\{d_i;\sqrt{K(\cdot, \beta_0 | \tilde{D})}, F_{\theta}\}$ are not bounded, and finally the influence function $\IF\{d_i;\sqrt{CLR(\cdot, \beta_0 | \tilde{D})}, F_{\theta}\}$ is not bounded. Therefore, we can conclude that the classical CLR statistic is not (locally) robust.

Intuitively, the classical CLR statistic inherits the robustness properties of the estimators it is constructed upon. As the LS estimators are not robust, the classical CLR statistic is also not robust. However, from Propositions \ref{prop-IF-RCLR} and \ref{prop-IF-tests} it follows that if the influence function of the estimators would be bounded, then the influence function of the general (robust) CLR statistic would be bounded. From Section~\ref{sec-m-estimator}, it follows that, when the function $\psi$ is bounded the influence function of the M-estimator is bounded. Therefore, the general (robust) CLR statistic introduced in (\ref{for-clr-stat}) is robust whenever the M-estimator it is constructed upon has a bounded (general) score function $\psi$. An example of an M-estimator with a bounded influence function is the M-estimator of Mallows type. In Section~\ref{sec:rob-clr}, we show how these types of estimators can be used to construct a robust CLR statistic that can be used in practice.

\subsection{Asymptotic Distributions} \label{sec-asymp-dist}
In this section, we derive the asymptotic distribution of the general (robust) AR, K and CLR statistics defined in (\ref{for:RAR}), (\ref{for:RK}) and (\ref{for-clr-stat}), and show how to construct tests based on these statistics. We start with a lemma that plays a key role in deriving all of the asymptotic distributions of the statistics.

\begin{lemma}\label{lemma-alternative-pi-est}
    Under the null hypothesis $\beta = \beta_0$, and the regularity conditions of Assumption~\ref{assumptions-main}, then as $n \rightarrow \infty$,
    \begin{align*}
        \sqrt{n}
       \left\{\begin{matrix}
            g(F_n, \beta_0) \\
            D(F_n, \beta_0)
        \end{matrix} \right\} 
            \overset{d}{\to} \mathcal{N}\left[\begin{pmatrix} 0 \\ \pi \end{pmatrix}, \left\{\begin{matrix}
                \Omega(F_{\theta}, \beta_0) & 0 \\ 0 & \Lambda(F_{\theta}, \beta_0)
            \end{matrix} \right\}
            \right].
    \end{align*}
\end{lemma}
\begin{proof}
    See Appendix.
\end{proof}
\noindent
From Lemma \ref{lemma-alternative-pi-est}, we see that under the null hypothesis $\beta = \beta_0$, as $n \rightarrow \infty$,
\begin{align}\label{for-test-RAR}
    nRAR(F_n, \beta_0) \overset{d}{\to} \chi_{k}^2.
\end{align}
 Moreover, under the null hypothesis $\beta = \beta_0$ and $\pi = 0$, as $n \rightarrow \infty$,
\begin{align*}
    nRW(F_n, \beta_0) \overset{d}{\to} \chi_{k}^2.
\end{align*}
Based on (\ref{for-test-RAR}), we can construct the general (robust) AR test as:
\begin{align*}
    \phi_{RAR}(\beta_0) = 1\{nRAR(F_n, \beta_0) > \chi_{k, 1-\alpha}^2 \},
\end{align*}
where $\chi_{k, 1-\alpha}^2$ denotes the $1-\alpha$ quantile of a $\chi_{k}^2$ distribution.

 Next, in Proposition \ref{prop-k-stat}, we derive the asymptotic distribution of the RK statistic.
\begin{proposition}\label{prop-k-stat}
    Under the null hypothesis $\beta = \beta_0$, and the regularity conditions in Assumption \ref{assumptions-main}, then as $n \rightarrow \infty$,
    \begin{align}
        nRK(F_n, \beta_0) \overset{d}{\to} \chi_1^2.
    \end{align}
\end{proposition}
\begin{proof}
    See Appendix.
\end{proof}
\noindent
Based on Proposition \ref{prop-k-stat}, we can introduce the general (robust) K test as:
\begin{align*}
    \phi_{RK}(\beta_0) = 1\{nRK(F_n, \beta_0) > \chi_{1, 1-\alpha}^2 \}.
\end{align*}

Finally, we derive the asymptotic distribution of the RCLR statistic.
\begin{proposition}\label{prop:rclr-dist}
Under the null hypothesis $\beta = \beta_0$, and the regularity conditions in Assumption \ref{assumptions-main}, it holds that, conditional on $D(F_n, \beta_0) = \tilde{D}$ and with $\tilde{W} = \tilde{D}^{\top}\Lambda(F_{\theta}, \beta_0)^{-1} \tilde{D}$, then as $n \rightarrow \infty$,
    \begin{align} \label{for-asymp-dist-clr}
        nRCLR(F_n, \beta_0) \overset{d}{\to} \frac{1}{2}\left\{\chi_{k - 1}^2 + \chi_{1}^2 - \tilde{W} + \sqrt{(\chi_{k - 1}^2 + \chi_{1}^2 + \tilde{W})^2 - 4 \tilde{W}\chi_{k - 1}^2 }\right\},
    \end{align}
    where $\chi_{k - 1}^2$ and $\chi_{1}^2$ are independent chi-squared distributed random variables with $k - 1$ and $1$ degrees of freedom.
\end{proposition}
\begin{proof}
    See Appendix.
\end{proof}
\noindent
Using Proposition~\ref{prop:rclr-dist} we can introduce the robust CLR test as follows:
\begin{align*}
    \phi_{RCLR}(\beta_0) = 1[nRCLR(F_n, \beta_0) > c_{\alpha}\{\sqrt{n}D(F_n, \beta_0)\} ],
\end{align*}
where $c_{\alpha}\{\sqrt{n}D(F_n, \beta_0)\}$ denotes the conditional $1-\alpha$ quantile of the asymptotic distribution given in (\ref{for-asymp-dist-clr}). The critical values and confidence sets can be computed using simulation and test inversion \citep{moreira2003conditional, andrews2019weak}. That is, we find all the values of $\beta_0$ for which the data does not reject the null hypothesis. The confidence set is then $\left\{\beta_0 \ |\ \beta_0 \in \mathbb{R} \text{ and } \phi_{RCLR}(\beta_0) = 0 \right\}$. 

\subsection{A Robust Conditional Likelihood Ratio Test of Mallows Type}\label{sec:rob-clr}
In this section, we show how to construct a robust CLR statistic based on Mallows type M-estimators. These estimators solve the following estimating equations:
\begin{align*}
    \frac{1}{n} \sum_{i=1}^n \omega(z_i) \rho\left\{\frac{y_{i} - z_{i}^{\top}\delta(F_n)}{\sigma_{\epsilon}(F_n)} \right\}z_i = 0, \\
    \frac{1}{n} \sum_{i=1}^n \omega(z_i) \rho\left\{\frac{x_{i} - z_{i}^{\top}\pi(F_n)}{\sigma_{v}(F_n)} \right\}z_i = 0,
\end{align*}
where we use $\omega(z_i) = \sqrt{1 - z_i^{\top}(Z^{\top}Z)^{-1}z_i}$ as a weight function, and $\rho \colon \mathbb{R} \mapsto \mathbb{R}$ is the Huber function:
\begin{align*}
    \rho(x) = 
                \begin{cases}
                    x,  &\text{ if } |x| \leq 1.345, \\ 
                    \text{sgn}(x) \cdot 1.345,  &\text{ if } |x| > 1.345. 
                \end{cases}
\end{align*}
The estimates $\sigma_{\epsilon}(F_n)$ and $\sigma_{v}(F_n)$ are robust estimates of the scales. In our case, for the practical implementation in Sections \ref{sec-simul-study} and \ref{sec:case-studies}, we use the function \texttt{rlm} from the R package \texttt{MASS} \citep{mass2002}. We use the default robust scale estimator provided, which is based on a re-scaled median absolute deviation of the residuals. In Sections \ref{sec-simul-study} and \ref{sec:case-studies}, for simplicity, we do the estimation equation per equation and not simulteneously. 

Now we only need to obtain estimates for the (co)variances belonging to these estimators. Using Equations (\ref{for-if-m-est-delta}) and (\ref{for-if-m-est-pi}), we derive the following influence functions for the Mallows type M-estimators:
\begin{align*}
    \IF\left\{d_i; \delta(\cdot), F_{\theta}\right\} &= \left\{\int \frac{w(z)}{\sigma_{\epsilon}} \rho'\left(\frac{y - z^{\top}\delta}{\sigma_{\epsilon}}\right)zz^{\top} d F_{\theta}\right\}^{-1}  \omega(z_i) \rho\left(\frac{y_i - z_i^{\top}\delta}{\sigma_{\epsilon}} \right)z_i, \\
    \IF\left\{d_i; \pi(\cdot), F_{\theta}\right\} &= \left\{\int \frac{w(z)}{\sigma_{v}} \rho'\left(\frac{x - z^{\top}\pi}{\sigma_v}\right)zz^{\top} d F_{\theta}\right\}^{-1}  \omega(z_i) \rho\left(\frac{x_i - z_i^{\top}\pi}{\sigma_{v}} \right)z_i,
\end{align*}
where 
\begin{align*}
    \rho'(x) = 
                \begin{cases}
                    1,  &\text{ if } |x| \leq 1.345, \\ 
                    0,  &\text{ if } |x| > 1.345. 
                \end{cases}
\end{align*}
Using these influence functions, we can calculate the (co)variance matrices. For instance, $\Sigma_{\delta \delta}(F_{\theta}) = \int \IF\{d; \delta(\cdot), F_{\theta}\}  \IF\{d; \delta(\cdot), F_{\theta}\}^{\top} dF_{\theta} = M_{\delta \delta}(F_{\theta})^{-1} Q_{\delta \delta}(F_{\theta}) M_{\delta \delta}(F_{\theta})^{-\top}$, where
\begin{align*}
    M_{\delta \delta}(F_{\theta}) &= \int \frac{w(z)}{\sigma_{\epsilon}} \rho'\left(\frac{y - z^{\top}\delta}{\sigma_{\epsilon}}\right)zz^{\top} d F_{\theta}, \\
    Q_{\delta \delta}(F_{\theta}) &= \int \omega(z)^2 \rho\left(\frac{y - z^{\top}\delta}{\sigma_{\epsilon}} \right)^2z z^{\top} dF_{\theta}.
\end{align*}
 Evaluating $M_{\delta \delta}(\cdot)$ and $Q_{\delta \delta}(\cdot)$ at the empirical distribution $F_n$ yields empirical estimate $\Sigma_{\delta \delta}(F_n)$ of the variance matrix $\Sigma_{\delta \delta}(F_{\theta})$. 

With these estimates, we can construct the robust CLR test (and also the robust AR and K tests). We study the performance of this test in different contaminated scenarios in Section \ref{sec-simul-study}. Note, this is just one possible way to construct a robust CLR test. Other promising estimators that can be used to construct a robust CLR statistic, are MM-estimators \citep{yohai1987high} and robust SUR estimators based on MM-estimators \citep{saraceno2021robust}.

\section{Simulation Study}\label{sec-simul-study}
In this section, we numerically evaluate the robustness properties of the robust CLR test introduced in Section \ref{sec:rob-clr} compared to the classical CLR test. We consider four different scenarios: a baseline scenario without contamination, two scenarios with contamination by a single outlier and scenario with ``distributional" contamination in the error terms. These contaminated scenarios are generated by slightly altering the baseline scenario. The exact details of these different contaminations are given below. We start by introducing the scenario without contamination. 

Let $\iota_3$ denote a 3-dimensional vector of ones. We generate data from the following model:
\begin{align}
x &= w +  \pi  z^{\top} \iota_3 + v, \\
y &=  \beta x + 2w + u.
\end{align}
 We generate three exogeneous instruments $z = (z_{1}, z_{2}, z_{3})^{\top}$ and one exogeneous control variable $w$. We draw $z_1, z_2, z_3$ and $w$ from independent standard normal distributions. The errors $u$ and $v$ are drawn from a bivariate normal distribution with variances equal to $1$ and correlation $\rho = 0.5$. We consider two different cases for the parameter $\pi \in \{0.1, 1\}$. When $\pi = 0.1$ we are in the weak instrument case and when $\pi = 1$ we are in the strong instrument case. The sample size is $n = 250$ and we repeat the study $10000$ times. In the simulation, we test $H_0 \colon \beta = 0$ at a 5\% significance level.

\subsection{Results}
\begin{figure}
    \centering
    \includegraphics[scale = 0.65]{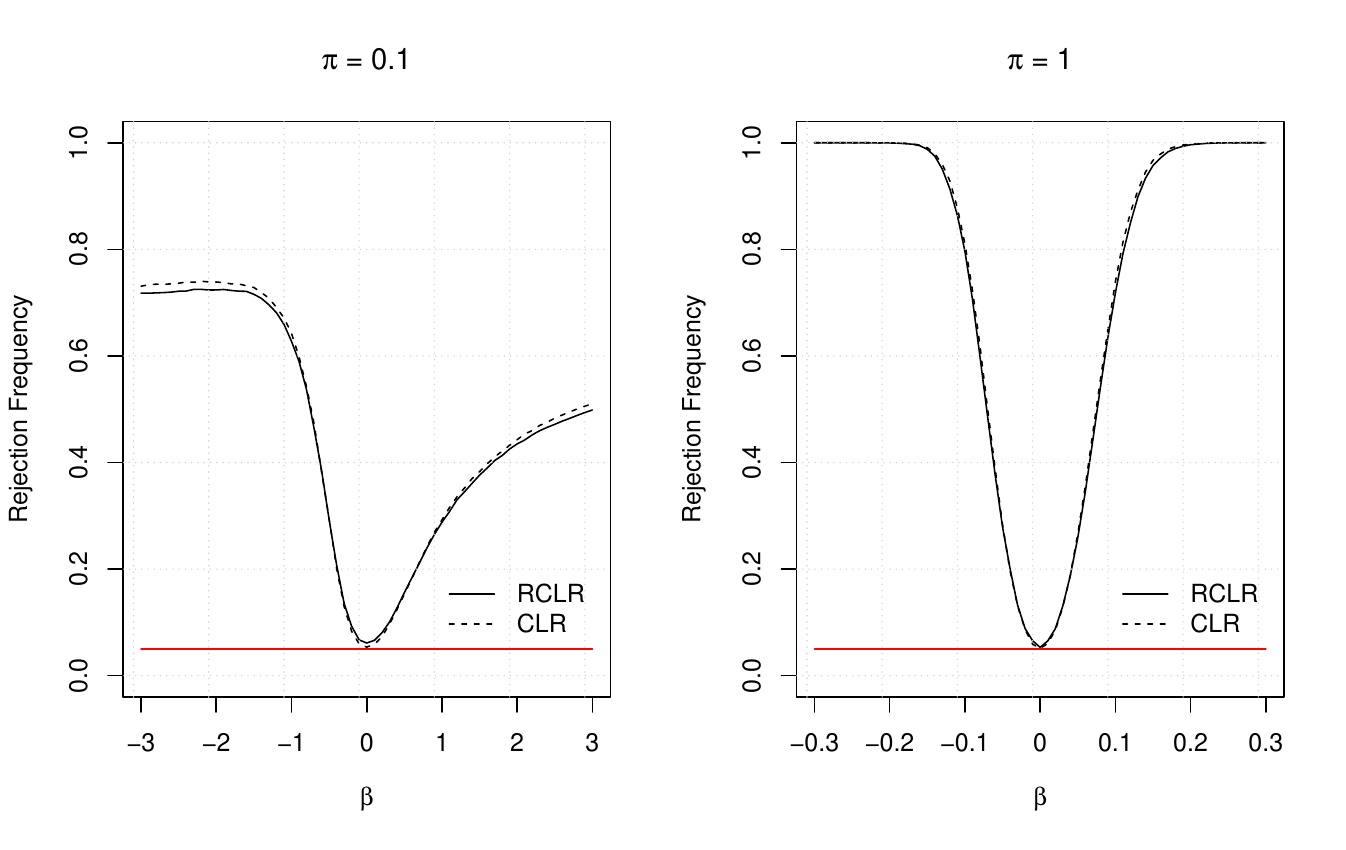}
    \caption{Power curves of robust CLR and CLR tests that test $H_0 \colon \beta = 0$ for various values of $\beta$ with three instruments, $\rho = 0.5$ and no contamination. }
    \label{fig-simulation-case1}
\end{figure}

In Figure \ref{fig-simulation-case1}, we show the power curves of the RCLR and CLR tests in a scenario without contamination. In this case, we see that both the RCLR and CLR tests are size correct in both the weak and strong instrument cases. Furthermore, we see that the CLR test is more powerful than the RCLR test in both the weak and strong instrument cases. However, we see that the loss of power of the RCLR test is not very large implying that the extra robustness comes at a low cost.

\begin{figure}
    \centering
    \includegraphics[scale = 0.65]{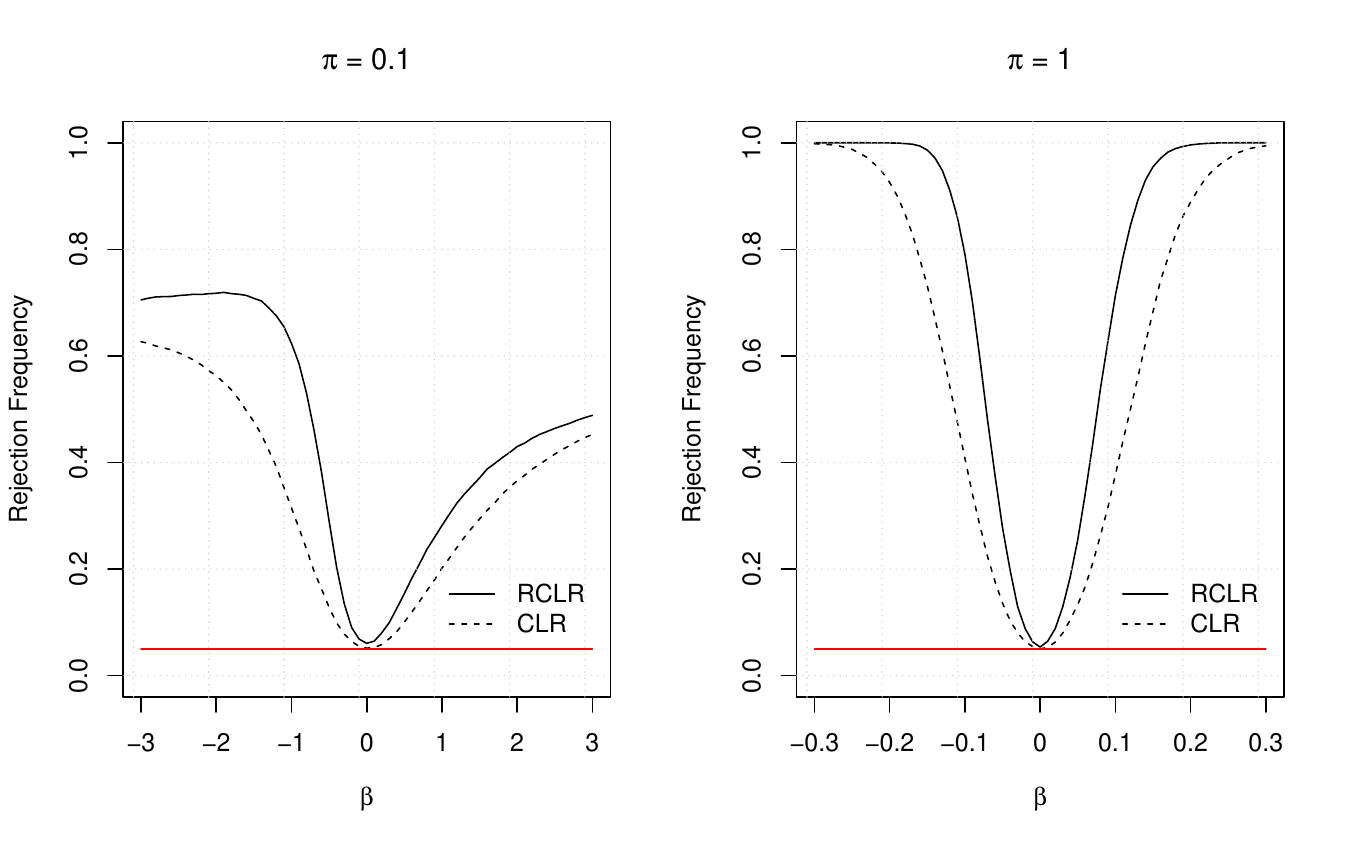}
    \caption{Power curves of robust CLR and CLR tests that test $H_0 \colon \beta = 0$ for various values of $\beta$ with three instruments, $\rho = 0.5$ and a large outlier in the $y$ variable.}
    \label{fig-simulation-case2}
\end{figure}

In Figure~\ref{fig-simulation-case2}, we show the power curves of the RCLR and CLR tests in a scenario with contamination by a large outlier in the $y$ variable. The data is first generated according to the baseline scenario and then an outlier is introduced. The outlier is generated by changing the first observation to $(y_{1}, x_{1}, z_{11}, z_{12}, z_{13}, w_{1}) = (20, x_{1}, z_{11}, z_{12}, z_{13}, w_{1})$, which means we only change the data point $y_{1}$ to $20$. In this case, we see that the RCLR test remains almost as powerful as in the case without contamination in both the weak and strong instrument cases. This happens, because the outlier is successfully downweighted by the robust M-estimators. In contrast, the CLR test loses power compared to the uncontaminated case and is now less powerful compared to the RCLR test. The loss of power happens, because the outlier increases the variance estimates of the classical test. This increased variance results in fewer rejections of the null, lowering the power. Overall, the CLR test remains reliable as we can see that it is still size correct. This happens because the (vertical) outlier does not bias the LS estimators $\delta(F_n)$ and $\pi(F_n)$. We would see a similar pattern if the outlier was in any other variable, for example the $x$ or $w$ variable, where the CLR test loses power but remains reliable. Therefore, we can ask the question whether it is necessary to use a robust CLR test. The simple answer is yes, as we demonstrate in the following contamination scenario. 

\begin{figure}
    \centering
    \includegraphics[scale = 0.65]{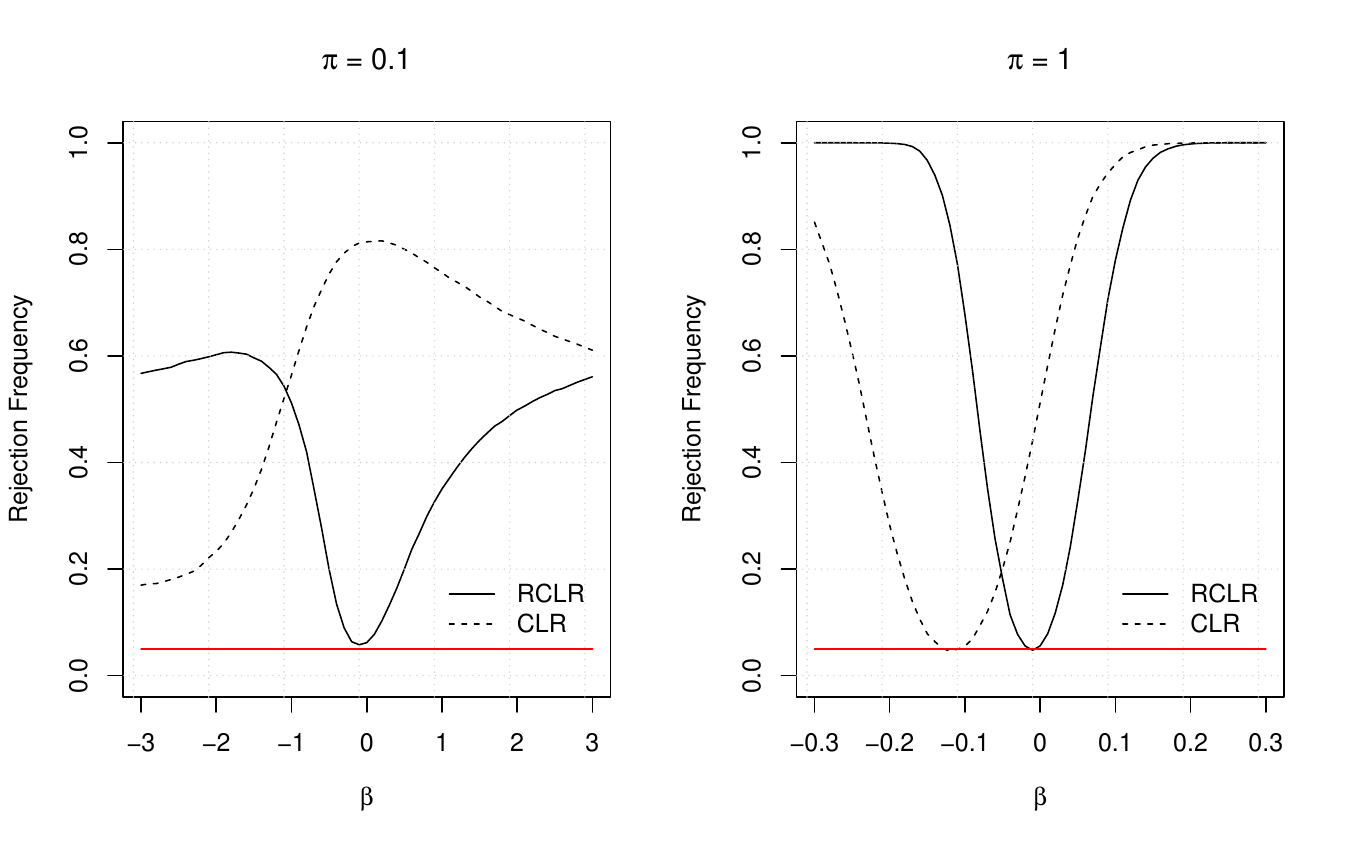}
    \caption{Power curves of robust CLR and CLR tests that test $H_0 \colon \beta = 0$ for various values of $\beta$ with three instruments, $\rho = 0.5$ and a large outlier in the $y$ variable combined with an outlier in the $z$ variable. }
    \label{fig-simulation-case3}
\end{figure}

In Figure \ref{fig-simulation-case3}, we show the power curves of the RCLR and CLR tests in a scenario with contamination by a large outlier in the $y$ and $z$ variable. This outlier was generated by replacing the first observation with $(y_{1}, x_{1}, z_{11}, z_{12}, z_{13}, w_{1}) = (20, x_{1}, 5, z_{12}, z_{13}, w_{1})$. We see that the power curves of the robust CLR test remain size correct in both weak and strong instrument cases and only lose a little bit of power compared to the scenario without contamination. This shows that the robust CLR test effectively downweights the outlier. However, this is not the case for the CLR test. We see that in both the strong and weak instrument cases the CLR test is not size correct anymore and the minimum is shifted to the left. To understand how this happens, for simplicity, we consider the case with only one instrument. In this case the CLR statistic behaves the same as the AR statistic. The AR statistic is minimized when $g(F_n) = \delta(F_n) - \pi(F_n) \beta_0 = 0$. The outlier we constructed can positively bias the estimator $\delta(F_n)$ when a nonrobust estimator is used. Hence, under the null, we obtain $\delta(F_n) - \pi(F_n) \beta_0 = \delta(F_n) \approx \delta + b = \pi \beta + b$, where $b > 0$ denotes a positive bias term. Solving $\pi \beta + b = 0$ for $\beta$ gives $\beta = -b/\pi$. This formula shows that if there is a positive bias $b$, then the minimum will not be at $\beta = 0$ implying there must an overrejection at $\beta = 0$. Furthermore, as in our case the bias is positive and $\pi > 0$, we obtain a negative value for $\beta$ so that the power curves shift to the left. At last, when the instrument is weak and $\pi \approx 0$, then the $\beta = -b/\pi$ term will be large showing that a (small) bias caused by outliers can be more problematic when the instrument is weak compared to the strong instrument case. This last finding also holds more generally for biases that are not necessarily due to outliers \citep{small2008war}.

\begin{figure}
    \centering
    \includegraphics[scale = 0.65]{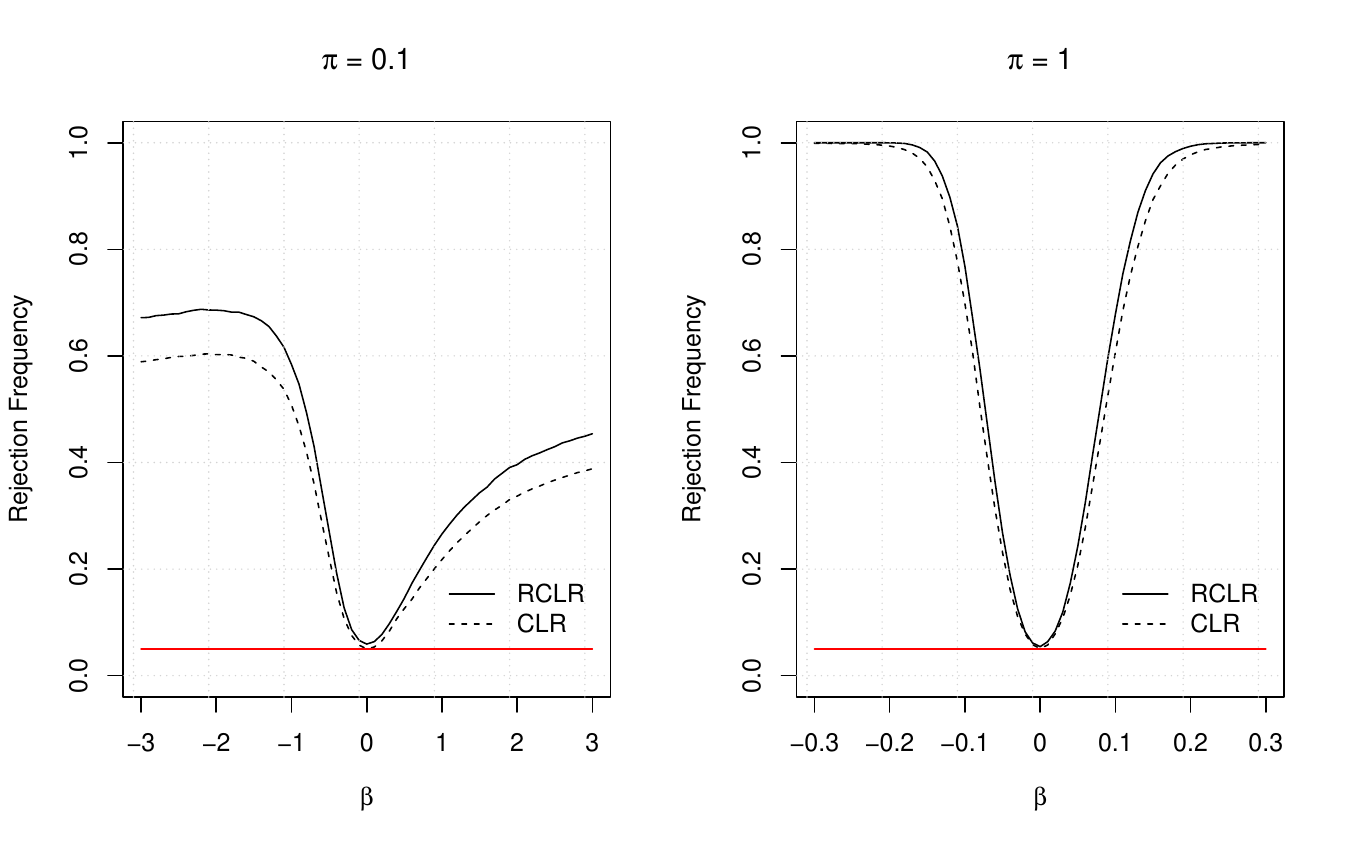}
    \caption{Power curves of robust CLR and CLR tests that test $H_0 \colon \beta = 0$ for various values of $\beta$ with three instruments, $\rho = 0.5$ and distributional contamination in the error term. }
    \label{fig-simulation-case4}
\end{figure}

In Figure \ref{fig-simulation-case4}, we show the power curves of the RCLR and CLR tests in a scenario with contamination in the error terms. In this case, the first $50$ error terms $(u_i, v_i), i = 1, \dots, 50$, are drawn from a mean zero bivariate $t(3)$-distribution with correlation $\rho = 0.5$ instead of a mean zero bivariate normal distribution. In this case, we see that the RCLR test is strictly more powerful than the CLR test in both the weak and strong instrument settings. This happens, because the error terms drawn from the $t(3)$-distribution sometimes introduce larger outlying values that increase the variance estimates the CLR statistic relies on. This results in fewer rejections of the null hypothesis when $\beta \neq \beta_0$ lowering the power. The RCLR effectively downweights these outlying values, resulting in better variance estimates, which results in a higher power. 

\section{Empirical Examples}\label{sec:case-studies}
In this section, we show how the robust CLR test introduced in Section \ref{sec:rob-clr} can be used in practice by revisiting three empirical studies. First, we consider the data and several specifications in \cite{alesina2011segregation} who examine the effect of segregation on the quality of government. Second, we revisit the main specifications considered in \cite{ananat2011wrong} where the effect of (racial) segregation on urban poverty and inequality is studied. Finally, we revisit the \cite{staiger1997instrumental} specifications for the \cite{angrist1991does} data where the effect of education on labor market earnings is studied.

\subsection{Alesina and Zhuravskaya (2011)}
\cite{alesina2011segregation} study the effect of segregation on the quality of government in a cross section of countries. They find that ethnically and linguistically segregated countries have a lower quality of government. Furthermore, they find that there is no relationship between religious segregation and governance. To address endogeneity concerns caused by mobility and endogeneous internal borders, an instrument is constructed for segregation. For more information, data and the construction of the instrument we refer to \cite{alesina2011segregation}.

In Section 5D \cite{alesina2011segregation} mention that they carefully examined whether a handful of influential observations drive their results. By exluding influential observations and recalculating their statistics they conclude that this is not the case. However, \cite{alesina2011segregation} mention that in the specifications of Panel~D in Table~7 removing two influential observations leads to the first-stage $F$-statistic dropping from 17.22 to 7.82 making inference based on 2SLS estimator unreliable \citep{staiger1997instrumental}. \cite{alesina2011segregation} solve this by also removing the most influential observation in the first stage from the data so that the instrument becomes ``strong enough". The practice of manually removing outliers from the data and then relying on classical statistical methods is not recommended for several reasons. The asymptotic distribution of the estimators and test statistics are unknown as they become conditional on the outlier detection procedure(s). The variability may be underestimated, which can lead to an overrejection of the null hypothesis. Lastly, some outliers might be difficult to detect due to masking effects. We refer to Section 4.3 in \cite{maronna2019robust} and Section 1.3 in \cite{heritier2009robust} for a more detailed discussion on these issues. Therefore, it is advisable to rely on robust estimation and testing procedures from the start. It seems that in case of \cite{alesina2011segregation} that the use of the 2SLS estimator might be questionable due to the outliers and the weak instrument. Therefore, we believe that it is desirable to re-evaluate the robustness of the results and the stability of the conclusions. We apply our test to six specifications (``Voice", ``Political stability", ``Government effectiveness", ``Regulatory Quality", ``Rule of law" and ``Control of corruption") of Panel D in Table 7 of \cite{alesina2011segregation}. As there is only one instrument, we calculate the $95\%$ confidence sets of the robust AR statistic and the (classical) AR statistic. The results are reported in Table \ref{tab:AZ2011}.

\begin{table}
\centering
\caption{Results using data from \cite{alesina2011segregation}. Specifications correspond to the specifications in Panel D of Table 7 in \cite{alesina2011segregation}. Confidence sets are given for the parameter belonging to endogeneous regressor $(x)$ ``Segregation" for six different specifications $(y)$. The RAR confidence sets are calculated based on Mallows Type estimators as in Section \ref{sec:rob-clr}.}
\label{tab:AZ2011}
\resizebox{\textwidth}{!}{%
\begin{tabular}{@{}lllllll@{}}
\toprule
                          & \multicolumn{6}{c}{Language}                          \\  \cline{2-7} 
       \makecell{Specification}             & \makecell{I} & \makecell{II} & \makecell{III} & \makecell{IV} & \makecell{V} & \makecell{VI} \\
                    & \makecell{Voice}            & \makecell{Political\\ stability}       & \makecell{Government \\ effectiveness}       & \makecell{Regulatory \\ quality}      & Rule of law     & \makecell{Control of \\ corruption}  \\ \midrule
$95\%$ AR confidence set                  & \makecell{$[-6.04,-0.93]$} & \makecell{$[-5.73, -1.01]$} & \makecell{$[-3.70, 0.79]$} & \makecell{$[-4.86, 2.20]$}  & \makecell{$[-3.62, 0.52]$} & \makecell{$[-3.48, 1.81]$} \\
$95\%$ RAR confidence set                  & 
\makecell{$[-9.67, -0.65]$}   & \makecell{$[-8.68, -0.63]$} & \makecell{$[-3.86, 1.37]$}   & \makecell{$[-2.40, 5.58]$} & \makecell{$[-3.31, 1.64]$}   & \makecell{$[-4.16,  2.21]$} \\
All control variables & \makecell{Yes}               & \makecell{Yes}               & \makecell{Yes}              & \makecell{Yes}              & \makecell{Yes}              & \makecell{Yes} \\
No. of observations & \makecell{92}               & \makecell{92}               & \makecell{92}              & \makecell{92}              & \makecell{92}             & \makecell{92} \\
First-stage $F$     & \makecell{$17.22$}         & \makecell{$17.22$}          & \makecell{$17.22$}         & \makecell{$17.22$}         & \makecell{$17.22$}         & \makecell{$17.22$}         \\\bottomrule
\end{tabular}%
}
\end{table}

From Table \ref{tab:AZ2011}, we can see that in specifications IV and V that the confidence set of the robust AR test is shifted compared to the AR confidence set. This suggests that outliers did have an effect as we would expect the confidence set of the robust AR test to be wider than the confidence set of the AR test when there are no outliers, but not shifted. The shift of the confidence set suggests that the LS estimators the AR test is constructed upon are biased due to the outlier(s). 
Therefore, the confidence sets based on the robust AR are more reliable. In specifications I, II, III and VI, we see that the robust AR confidence sets are wider, but not (much) shifted, compared to the AR confidence set. This suggests that outliers did not have a large effect. Overall, the outliers do not seem to be very problematic as in all specifications the final decision whether to reject or not reject the null hypothesis $H_0 \colon \beta = 0$ remains the same.

\subsection{Ananat (2011)}
\cite{ananat2011wrong} studies the effect of racial segregation on urban poverty and inequality. To overcome endogeneity issues, a railroad division index is used to instrument for racial segregation. Using this instrumental variable, \cite{ananat2011wrong} shows that segregation increases metropolitan rates of black poverty and overall black-white income disparities, while decreasing rates of white poverty and inequality within the white population. For more information, data and the construction of the instrument we refer to \cite{ananat2011wrong}.

\cite{klooster2021outlier} show that an outlier in the control variable used in the main results of \cite{ananat2011wrong} inflates the first-stage $F$-statistic from 1.83 to 19.32. As the outlier was not taken into account in the original study, it was assumed that the instrument was strong. Consequently, estimation was done with a 2SLS estimator and inference with a $t$-test. Due to the outlier (and the weak instrument) inference based on the 2SLS estimator might be unreliable. Therefore, to re-evaluate the robustness of the results, we apply our test to the four main specifications (``Gini index whites", ``Gini index blacks", ``Poverty rate whites", ``Poverty rate blacks"), which can be found in columns (3) and (4) of Table 2 in \cite{ananat2011wrong}. As there is only one instrument, we calculate the $95\%$ confidence sets of the robust AR statistic and the (classical) AR statistic. The results are reported in Table~\ref{tab:ANANAT2011}.

\begin{table}
\centering
\caption{Results using data from \cite{ananat2011wrong}. Specifications correspond to the specifications in columns (3) and (4) in Table 2 in \cite{ananat2011wrong}. Confidence sets are given for the parameter belonging to endogeneous regressor $(x)$ ``Segregation" for four different specifications $(y)$. The RAR confidence sets are calculated based on Mallows Type estimators as in Section \ref{sec:rob-clr}.}
\label{tab:ANANAT2011}
\begin{tabular}{@{}lllll@{}}
\toprule
Specification            & \makecell{I}               &  \makecell{II}                 &        \makecell{III}         &     \makecell{IV}            \\ & \makecell{Gini index \\ whites} & \makecell{Gini index \\ blacks} & \makecell{Poverty rate \\ whites} & \makecell{Poverty rate \\ blacks}
       \\  \midrule
95\% AR confidence set &  \makecell{$(-0.64, -0.18)$} & \makecell{$(0.22, 2.15)$}  & \makecell{$(-0.38, -0.09)$} & \makecell{$(0.00, 0.48)$}  \\ 
95\% RAR confidence set  & \makecell{$(-\infty, -0.15)$ \\ \quad $\cup (2.60, \infty)$}   &  \makecell{$(-\infty, -6.21)$ \\ \quad $\cup (0.23, \infty)$}     & \makecell{$(-\infty, -0.09)$ \\ \quad  $\cup (1.45, \infty)$}    &    \makecell{$(-\infty, 0.77)$ \\ \quad  $\cup (2.30, \infty)$}             \\
First-stage $F$     & \makecell{$19.32$}         & \makecell{$19.32$}          & \makecell{$19.32$}         & \makecell{$19.32$} \\
No. of observations & \makecell{121}               & \makecell{121}               & \makecell{121}              & \makecell{121}                           \\  \bottomrule
\end{tabular}
\end{table}

When we analyze Table \ref{tab:ANANAT2011}, we find large differences between the robust and classical confidence sets. When there are no outliers in the data, we would expect that the robust confidence sets are only a bit wider than the classical confidence sets. However, in this case, the classical confidence sets are bounded convex sets, while the robust confidence sets are unbounded sets. The shape of the robust confidence sets do correctly suggest that the instrument is weak. In this example, the outlier does have a large effect and we recommend using the robust confidence sets for reliable inference.

\subsection{Angrist and Krueger (1991)}
\cite{angrist1991does} study the effect of education on labor market earnings. To adress endogeneity issues, quarter of birth instruments are constructed for education. Using these instrumental variables they find a positive relationship between years of education and labor market earnings. We revisit four specifications presented in Table 2 of \cite{staiger1997instrumental} based on the 1930 - 1939 cohort. For more information, data and the construction of the instrument, we refer to  \cite{staiger1997instrumental} and \cite{angrist1991does}.

 \cite{BoundJaegerBaker1995} showed that the relationship between the instruments and the endogeneous regressor is quite weak in certain specifications of \cite{angrist1991does}. Furthermore, more recently, \cite{solvsten2020robust} shows that the LIML residuals of the structural equation of a certain specification in \cite{angrist1991does} is distributed roughly like a normal distribution at the center with outlying errors that closely follow a $t(3)$-distribution (reminiscent of the ``distributional" contamination scenario in Section \ref{sec-simul-study}). Due to the weak instruments (and possible outliers), inference based on the 2SLS estimator used by \cite{angrist1991does} might be unreliable. Moreover, due to outliers, weak instrument robust tests might be corrupted and/or inefficient. Therefore, to re-evaluate the robustness of the results we replicate the results reported in Panel A of Table 2 in \cite{staiger1997instrumental}. For each specification, we give the $95\%$ confidence set of the CLR and robust CLR, and the first-stage $F$-statistic. The results are reported in Table \ref{tab:AK1991}.
 
 When we compare the $95\%$ confidence sets of the CLR and RCLR statistics, we note that the RCLR confidence sets are smaller than the CLR confidence sets in every specification. This happens because the RCLR statistic effectively downweights outlying values in the residuals. Similar as in the ``distributional" contamination scenario in Section~\ref{sec-simul-study} this results in better variance estimates and hence tighter confidence sets.

\begin{table}
\centering
\caption{Results for \cite{angrist1991does} data. Specifications as in Table 2 of \cite{staiger1997instrumental}, except for specification III (see the note below). Confidence sets are given for the parameter belonging to the endogeneous regressor $(x)$ ``Years of schooling" for four different specifications that all use the same dependent variable $(y)$ ``log weekly wages". The RCLR confidence sets are calculated based on Mallows type estimators as in Section \ref{sec:rob-clr}. QOB, YOB and SOB stand for quarter, year and state of birth.} 
\label{tab:AK1991}
\begin{tabular}{@{}lllll@{}}
\toprule
Specification                      & \makecell{I}                & \makecell{II}                & \makecell{III$^*$}                & \makecell{IV}                \\  \midrule
$95\%$ CLR confidence set & \makecell{ $[0.042, 0.136]$} & \makecell{$[0.026, 0.116]$}  & $[-0.064, 0.278]$ & $[-0.068, 0.265]$  \\
$95\%$ RCLR confidence set                 & $[0.047, 0.122]$   & $[0.032, 0.100]$    & $[-0.038, 0.189]$    &    $[-0.047, 0.179]$               \\
First-stage F        & \makecell{$30.53$}          & \makecell{$4.74$}            & \makecell{$2.43$}            & \makecell{$1.87$}            \\ 
\textit{controls $(w)$}    &                  &                   &                   &                   \\ 
Base controls        & \makecell{Yes}              & \makecell{Yes}               & \makecell{Yes}               & \makecell{Yes}               \\
SOB                  & \makecell{No}               & \makecell{No}                & \makecell{Yes}                & \makecell{Yes}               \\
Age, Age$^2$         & \makecell{No}               & \makecell{No}                & \makecell{No}               & \makecell{Yes}               \\
\textit{Instruments $(z)$} &                  &                   &                   &                   \\
QOB                  & \makecell{Yes}              & \makecell{Yes}               & \makecell{Yes}               & \makecell{Yes}               \\
QOB*YOB              & \makecell{No}               & \makecell{Yes}               & \makecell{Yes}               & \makecell{Yes}               \\
QOB*SOB              & \makecell{No}               & \makecell{No}                & \makecell{Yes}                & \makecell{Yes}              \\
No. of instruments   & \makecell{3}                & \makecell{30}                & \makecell{180}                & \makecell{178}               \\
Observations         & \makecell{329,509}          & \makecell{329,509}           & \makecell{329,509}           & \makecell{329,509}            \\ \bottomrule
\end{tabular}
    \begin{tablenotes}
      \small
      \item *This specification slightly different from specification III of Table 2 in \cite{staiger1997instrumental}. Instead of using Age and Age$^2$ as control variables, we use the SOB controls instead. This was done as we encountered some small numerical difficulties when replicating the original specification leading to unusual confidence sets for both the CLR and RCLR tests.
    \end{tablenotes}
\end{table}

\section{Conclusion}\label{sec-conclusion}
In this article, we proposed a general framework to construct weak instrument robust testing procedures that are also robust to outliers in the linear instrumental variable model. The framework is constructed upon M-estimators and we showed that the classical weak instrument robust tests, such as the AR, K and CLR tests, can be obtained by specifying the M-estimators to be the LS estimators. We formally showed that influence functions of the classical test statistics are not bounded and hence not (locally) robust. As the classical test statistics are not robust, we showed how to construct a robust CLR test based on a Mallows type M-estimator. By means of a simulation study, we documented good performance of our robust CLR test in different contamination scenarios. Finally, we illustrated how the robust CLR test can be used in practice by revisiting three different empirical studies.

\newpage
\section*{Appendix}

\begin{assumption}[Adapted from \cite{heritier1994robust}]
\label{assumptions-main}
Let $F$ be any arbitrary distribution on $\mathbb{R}^{2(p + k)}$ and define 
\begin{align*}
    K_F(\theta) &= \int \Psi(d, \theta) dF,
\end{align*}
with $\Psi(d, \theta) = \begin{bmatrix} \psi\{(y,z), \delta\}^{\top} & \psi\{(x,z), \pi\}^{\top} \end{bmatrix}^{\top}$.
\begin{enumerate}
    \item $K_F(\theta)$ exists at least on a (nondegenerate) open set $\mathcal{O}$.
    \item There exists $\theta^* \in \mathcal{O}$ satisfying $K_F(\theta^*) = 0$.
    \item  $\int \Psi(d, \theta) d F_{\theta} = 0$ (Fisher consistency).
    \item $\Psi(d, \theta)$ is a $2(p + k) \times 1$ vector function that is continuous and bounded on $\mathcal{D} \times \Theta$, where $\Theta$ is some nondegenerate compact interval containing $\theta^*$.
    \item $\Psi(d, \theta)$ is locally Lipschitz in $\theta$ about $\theta^*$, i.e., there exists a constant $\alpha$ such that 
    \begin{align*}
        ||\Psi(d, \theta) - \Psi(d, \theta^*)|| \leq \alpha ||\theta - \theta^*||
    \end{align*}
    uniformly in $d \in \mathcal{D}$ and for all $\theta$ in a neighbourhood of $\theta^*$.
    \item The generalized Jacobian $\partial K_F(\theta)$ is of maximal rank at $\theta = \theta^*$.
    \item Given $\Delta > 0$, there exists $\epsilon > 0$ such that for all distributions in a $\epsilon$ neighborhood of $F$, $\sup_{\theta \in \Theta}||K_G(\theta) - K_F(\theta)|| > \delta$ and $\partial K_G(\theta) \subset \partial K_F(\theta) + \Delta B$, uniformly in $\theta \in \Theta$, where $B$ is the unit ball of $2(p + k) \times 2(p + k)$ matrices.
    \item $K_F(\theta)$ has at least a continuous derivative $(\partial/\partial \theta) K_F(\theta)$ at $\theta = \theta^*$.
    \item $(\partial \Psi/\partial \theta)(d,\theta)$ exists for $\theta = \theta^*$ and almost everywhere in a neighborhood of $\theta^*$. Moreover, for all $\theta$ in this neighborhood, there exists an integrable function $v(d)$ such that $||(\partial \Psi/\partial \theta)(d,\theta) || \leq v(d)$ a.e.
\end{enumerate}

\end{assumption}

\subsection*{Proof of Lemma \ref{lemma-alternative-pi-est}}
\begin{proof}
Assumptions 1.4 - 1.8 ensure Fr\'{e}chet differentiability \citep{clarke1986nonsmooth}, so that we can write
\begin{align*}
    T(F_n) - \theta = \int \IF\left\{d, T(\cdot), F_{\theta}\right\} dF_{\theta} + \mathsf{o}_{\mathsf{p}}(||F_n - F_{\theta}||_{\infty}).
\end{align*}
Therefore, we have
\begin{align*}
    \sqrt{n}
        \begin{Bmatrix}
            \delta(F_n) - \delta\\
            \pi(F_n) - \pi
        \end{Bmatrix}
        = \frac{1}{\sqrt{n}} \sum_{i=1}^n
        \begin{bmatrix}
             \IF\{d_i; \delta(\cdot), F_{\theta}\} \\
             \IF\{d_i; \pi(\cdot), F_{\theta}\}
        \end{bmatrix}
        + \mathsf{o}_{\mathsf{p}}(\sqrt{n}||F_n - F_{\theta}||_{\infty}).
\end{align*}
We can use the fact that $\sqrt{n}||F_n - F||_{\infty} = \mathsf{O}_{\mathsf{p}}(1)$ \citep{dvoretzky1956asymptotic}, and apply the Lindeberg-Feller Theorem to obtain, as $n \rightarrow \infty$,
\begin{align*}
        \sqrt{n}
        \begin{Bmatrix}
            \delta(F_n) \\
            \pi(F_n) 
        \end{Bmatrix}
        \overset{d}{\to} 
            \mathcal{N} 
                \left[ 
                    \begin{pmatrix}
                        \delta \\ 
                        \pi 
                    \end{pmatrix}, 
                    \begin{Bmatrix} 
                        \Sigma_{\delta \delta}(F_{\theta}) & \Sigma_{\delta \pi}(F_{\theta}) \\
                        \Sigma_{\pi \delta}(F_{\theta}) & \Sigma_{\pi \pi }(F_{\theta})
                    \end{Bmatrix} 
                \right].
\end{align*}
Note,
\begin{align*}
    \begin{bmatrix} 
        I_{k} &  -\left\{\Sigma_{\pi \delta}(F_{\theta}) - \Sigma_{\pi \pi}(F_{\theta}) \beta_0\right\}\Omega(F_{\theta},\beta_0)^{-1} \\
        -\beta_0 I_{k} & \left\{\Sigma_{\pi \delta}(F_{\theta}) - \Sigma_{\pi \pi}(F_{\theta}) \beta_0\right\}\Omega(F_{\theta}, \beta_0)^{-1}\beta_0 + I_{k}
    \end{bmatrix}^{\top}
           \begin{Bmatrix}
            \delta(F_n) \\
            \pi(F_n) 
        \end{Bmatrix}
        =
        \begin{Bmatrix}
            g(F_n) \\
            D(F_n)
        \end{Bmatrix},
\end{align*}
we conclude that, as $n \rightarrow \infty$,
\begin{align*}
    \sqrt{n}
        \begin{Bmatrix}
            g(F_n, \beta_0) \\
            D(F_n, \beta_0)
        \end{Bmatrix}
            \overset{d}{\to} \mathcal{N}\left[\begin{pmatrix} 0 \\ \pi \end{pmatrix}, \begin{Bmatrix}
                \Omega(F_{\theta}, \beta_0) & 0 \\ 
                0 & \Lambda(F_{\theta}, \beta_0)
            \end{Bmatrix}
            \right].
\end{align*}
\end{proof}

\subsection*{Proof of Proposition \ref{prop-IF-RCLR}}
\begin{proof}
For simplicity, in this proof we surpress the dependency on $\beta_0$ and $\tilde{D}$ in all the test statistics. Let $G$ be a general distribution function and define $\tilde{W} = \tilde{D}^{\top}\Lambda(F_{\theta}, \beta_0)^{-1}\tilde{D}$. The functional form of the RCLR statistic, conditional on $\tilde{D}$, is
\begin{align*}
    RCLR(G) = \frac{1}{2}\left[ RAR(G) - \tilde{W} + \sqrt{\left\{ RAR(G) - \tilde{W}\right\}^2 + 4\tilde{W} \cdot RK(G) } \right].
\end{align*}
To simplify the notation, we write
\begin{align*}
    A(G) = \left\{ RAR(G) - \tilde{W}\right\}^2 + 4\tilde{W} \cdot RK(G),
\end{align*}
so that
\begin{align*}
    RCLR(G) = \frac{1}{2}\left\{ RAR(G) - \tilde{W} + \sqrt{A(G)} \right\}.
\end{align*}

We start with the case $\tilde{D} \neq 0$. In this case it must hold that $\tilde{W} > 0$ and as $A(F_{\theta}) = \tilde{W}^2$, we have $RCLR(F_{\theta}) = 0$. To calculate the first derivative, note that
\begin{align*}
    \frac{\partial}{\partial t} RCLR(F_t)\Big|_{t=0} = \frac{1}{2} \left\{ \frac{\partial}{\partial t}RAR(F_t)\Big|_{t=0} + \frac{1}{2\sqrt{A(F_{\theta})}}\cdot \frac{\partial}{\partial t} A(F_t)\Big|_{t=0}  \right\}.
\end{align*}
We know that $\frac{\partial}{\partial t}RAR(F_t)\big|_{t=0}$ = 0, and 
\begin{align*}
    \frac{\partial}{\partial t} A(F_t)\Big|_{t=0} &= 2\left\{RAR(F_{\theta}) - \tilde{W}\right\} \frac{\partial}{\partial t}RAR(F_t)\Big|_{t=0} + 4\tilde{W} \frac{\partial}{\partial t}RK(F_t)\Big|_{t=0}\\
    &= 0,
\end{align*}
as $\frac{\partial}{\partial t}RAR(F_t)\big|_{t=0} = 0$, and $ \frac{\partial}{\partial t}RK(F_t)\big|_{t=0} = 0$.
Furthermore, as $\tilde{W} > 0$, we have
\begin{align*}
    A(F_{\theta}) = \{RAR(F_{\theta}) - \tilde{W}\}^2 + 4\tilde{W}RK(F_{\theta}) = (0 - \tilde{W})^2 + 0 = \tilde{W}^2.
\end{align*}
Hence, $A(F_{\theta}) > 0$ so that we are not dividing by zero. It thus follows that
\begin{align*}
    \frac{\partial}{\partial t} RCLR(F_t)\Big|_{t=0} = 0.
\end{align*}
Next, we calculate the second derivative. We have
\begin{align*}
    \frac{\partial^2}{\partial t^2} RCLR(F_t)\Big|_{t=0} =  \frac{1}{2}\left\{ \frac{\partial^2}{\partial t^2} RAR(F_t)\Big|_{t=0} + \frac{1}{2\tilde{W}} \frac{\partial^2}{\partial t^2} A(F_t)\Big|_{t=0} \right\},
\end{align*}
where we used the results that $\frac{\partial}{\partial t} A(F_t)\big|_{t=0} = 0$ and $A(F_{\theta}) = \tilde{W}^2$. We continue and obtain
\begin{align*}
    \frac{\partial^2}{\partial t^2} A(F_t)\Big|_{t=0} &= 2 \left\{\frac{\partial }{\partial t} RAR(F_t)\Big|_{t=0}\right\}^2 - 2\tilde{W}\frac{\partial^2 }{\partial t^2} RAR(F_t)\Big|_{t=0} + 4\tilde{W} \frac{\partial^2 }{\partial t^2} RK(F_t)\Big|_{t=0}\\
    &= - 2\tilde{W} \frac{\partial^2 }{\partial t^2} RAR(F_t)\Big|_{t=0} + 4\tilde{W} \frac{\partial^2 }{\partial t^2} RK(F_t)\Big|_{t=0}.
\end{align*}
Substituting this back into the previous equation, it follows that
\begin{align*}
    \frac{\partial^2}{\partial t^2} RCLR(F_t)\Big|_{t=0} &= \frac{1}{2}\left\{ \frac{\partial^2}{\partial t^2} RAR(F_t)\Big|_{t=0} + \frac{4\tilde{W} \frac{\partial^2 }{\partial t^2} RK(F_t)\Big|_{t=0} - 2\tilde{W} \frac{\partial^2 }{\partial t^2} RAR(F_t)\Big|_{t=0}}{2\tilde{W}}  \right\} \\
    &= \frac{\partial^2 }{\partial t^2} RK(F_t)\Big|_{t=0}.
\end{align*}
Therefore, using L'H\^{o}pital's rule twice, the influence function of the CLR statistic, given $\tilde{W} > 0$, is
\begin{align*}
    \IF\{d_i; \sqrt{RCLR}, F\} &= \lim_{t \rightarrow 0} \left\{\sqrt{RCLR(F_t)} - \sqrt{RCLR(F)}\right\}/t \\
    &=\left\{\lim_{t \rightarrow 0} RCLR(F_t)/t^2\right\}^{1/2} \\
    &= \left\{ \frac{1}{2}\frac{\partial^2}{\partial t^2} RCLR(F_t) \Big|_{t=0} \right\}^{1/2} \\
    &= \left\{ \frac{1}{2}\frac{\partial^2}{\partial t^2} RK(F_t) \Big|_{t=0} \right\}^{1/2}\\
    &= \IF(d_i; \sqrt{RK}, F).
\end{align*}

Next, we assume $\tilde{D} = 0$. In this case $\tilde{W} = 0$, so that
\begin{align*}
    RCLR(G) = RAR(G).
\end{align*}
Hence,
\begin{align*}
    \IF(d_i; \sqrt{RCLR}, F)  = \IF(d_i; \sqrt{RAR},F).
\end{align*}
\end{proof}

\subsection*{Proof of Proposition \ref{prop-IF-tests}}
\begin{proof}
To simplify the notation, we suppress the dependency on $\beta_0$ for the RAR and RK statistics. We have $\frac{\partial}{\partial t} RAR(F_t)\big|_{t=0} = 2 \IF\{d_i; g(\cdot, \beta_0), F_{\theta}\}^{\top}\Omega(F_{\theta}, \beta_0)^{-1} g(F_{\theta}, \beta_0) = 0$, as $g(F_{\theta}, \beta_0) = 0$ due to Fisher consistency. The second derivative gives
\begin{align*}
    \frac{\partial^2}{\partial t^2} RAR(F_t)\Big|_{t=0} = 2 \IF\{d; g(\cdot, \beta_0), F_{\theta}\}^{\top} \Omega(\beta_0)^{-1} \IF\{d; g(\cdot, \beta_0\}, F_{\theta}).
\end{align*}
Using L'H\^{o}pital's rule twice, we obtain
\begin{align*}
    \IF(d_i; \sqrt{RAR}, F_{\theta})  &= \lim_{t \rightarrow 0} \left\{\sqrt{RAR(F_t)} - \sqrt{RAR(F)}\right\}/t \\
    &=\left\{\lim_{t \rightarrow 0} RAR(F_t)/t^2\right\}^{1/2} \\
    &= \left\{ \frac{1}{2}\frac{\partial^2}{\partial t^2} RAR(F_t) \big|_{t=0} \right\}^{1/2}\\
    &= \sqrt{\IF\{d_i; g(\cdot, \beta_0), F_{\theta}\}^{\top} \Omega(F_{\theta}, \beta_0)^{-1}\IF\{d_i; g(\cdot, \beta_0), F_{\theta}\}}.
\end{align*}

Similarly, for the RK statistic we have
\begin{align*}
    \frac{\partial }{\partial t} RK(F_t) \Big|_{t=0} =\ &2 \IF\{d_i; g(\cdot, \beta_0), F_{\theta}\}^{\top}\pi\left\{\pi^{\top} \Omega(F_{\theta}, \beta_0) \pi \right\}^{-1}\pi^{\top} g(F_{\theta}, \beta_0)\\
    =&\ 0
\end{align*}
as $g(F_{\theta}, \beta_0) = 0$. The second derivative gives
\begin{align*}
    \frac{\partial^2 }{\partial t^2} RK(F_t)\Big|_{t=0} = 2 \IF\{d_i; g(\cdot, \beta_0), F_{\theta}\}^{\top}\pi\left\{\pi^{\top} \Omega(\beta_0) \pi \right\}^{-1}\pi^{\top} \IF\{d_i; g(\cdot, \beta_0), F_{\theta}\}.
\end{align*}
Again, using L'Hopitals Rule twice, we obtain
\begin{align*}
    \IF(d_i; \sqrt{RK}, F)  &= \lim_{t \rightarrow 0} \left\{\sqrt{RK(F_t)} - \sqrt{RK(F)}\right\}/t \\
    &=\left\{\lim_{t \rightarrow 0} RK(F_t)/t^2\right\}^{1/2} \\
    &= \left\{ \frac{1}{2}\frac{\partial^2}{\partial t^2} RK(F_t) \big|_{t=0} \right\}^{1/2}\\
    &= \sqrt{\IF\{d_i; g(\cdot, \beta_0), F_{\theta}\}^{\top}\pi\left\{\pi^{\top} \Omega(\beta_0) \pi \right\}^{-1}\pi^{\top} \IF\{d_i; g(\cdot, \beta_0), F_{\theta}\}}
\end{align*}

We omit the proof of the IF of the RW statistic, as it follows the exact same steps as the IF of the RAR statistic. We also omit the proof of the influence function of the K statistic, conditional on $D(F_n, \beta_0) = \tilde{D}$, as this also follows the exact same steps as the IF of the RAR statistic.
\end{proof}

\subsection*{Proof of Proposition \ref{prop-k-stat}}
\begin{proof}
     In this proof we follow similar arguments as in Proof 1 of \cite{kleibergen2005testing}. From Lemma \ref{lemma-alternative-pi-est}, we know that, as $n \rightarrow \infty$,
    \begin{align*}
        \sqrt{n}g(F_n, \beta_0) \overset{d}{\to} N \sim \mathcal{N}\{0, \Omega(F_{\theta}, \beta_0)\}.
    \end{align*}
     We denote the asymptotic distribution of $D(F_n, \beta_0)$ by $D$, i.e., $\sqrt{n} D(F_n, \beta_0)  \overset{d}{\to} D \sim \mathcal{N}\{\pi, \Lambda(\beta_0)\}$. Then, as $n \rightarrow \infty$, $\sqrt{n} D(F_n, \beta_0)^{\top} \sqrt{n}g(F_n, \beta_0) \overset{d}{\to} D^{\top} N$. The conditional distribution of $D^{\top} N$ given $D$ reads 
\begin{align*}
    D^{\top} N \big| D \sim \mathcal{N}\{0, D^{\top} \Omega(F_{\theta}, \beta_0) D\}.
\end{align*}
From Lemma \ref{lemma-alternative-pi-est}, we know that $D$ is independent of $N$ as they are jointly normally distributed and uncorrelated. Therefore, we obtain an unconditional result by normalizing the expression by $\{D^{\top}\Omega(F_{\theta}, \beta_0)D\}^{-1/2}$. Thus, as $n \rightarrow \infty$,
\begin{align*}
    &\{D(F_n, \beta_0)^{\top}\Omega(F_{\theta}, \beta_0)D(F_n, \beta_0)\}^{-1/2}D(F_n, \beta_0) \sqrt{n}g(F_n, \beta_0) \\
    & \quad \quad \quad \quad  = \{\sqrt{n}D(F_n, \beta_0)^{\top}\Omega(F_{\theta}, \beta_0) \sqrt{n}D(F_n, \beta_0)\}^{-1/2}\sqrt{n}D(F_n, \beta_0) \sqrt{n}g(F_n, \beta_0) \\
    & \quad \quad \quad \quad\overset{d}{\to} \{D^{\top}\Omega(F_{\theta}, \beta_0) D\}^{-1/2} D^{\top} N \sim \mathcal{N}(0, 1).
\end{align*}
Hence, as $n \rightarrow \infty$, $nRK(F_n, \beta_0) \overset{d}{\to} \chi_1^2$.
\end{proof}

\subsection*{Proof of Proposition \ref{prop:rclr-dist}}
\begin{proof}
Define $U(F_n) = \Omega(F_{\theta}, \beta_0)^{-1/2} g(F_n, \beta_0)$ and $R(F_n) = \Omega(F_{\theta}, \beta_0)^{1/2}D(F_n, \beta_0)$. Note, by Lemma \ref{lemma-alternative-pi-est}, as $n \rightarrow \infty$, $\sqrt{n}\begin{Bmatrix} U(F_n) & R(F_n) \end{Bmatrix}^{\top} \overset{p}{\rightarrow} \begin{pmatrix} U & R \end{pmatrix}^{\top}$, where $U$ and $R$ are independent normal distributions. 

We have $RAR(F_n, \beta_0) = U(F_n)^{\top}U(F_n)$ and $RK(F_n, \beta_0) = U(F_n)^{\top}P_{R(F_n)}U(F_n)$, where $P_{R(F_n)} = R(F_n)\left\{R(F_n)^{\top} R(F_n)\right\}^{-1} R(F_n)^{\top}$.
We can write $RAR(F_n, \beta_0) = RK(F_n, \beta_0) + RJ(F_n, \beta_0)$, with
\begin{align*}
    RJ(F_n, \beta_0) = U(F_n)^{\top}\left\{I_{k} - P_{R(F_n)}\right\}U(F_n).
\end{align*}
It holds that $\left\{I_{k} - P_{R(F_n)}\right\}^{-1/2}\sqrt{n}U(F_n) \overset{p}{\to} \left(I_{k} - P_{R}\right)U$, with $U \sim \mathcal{N}(0, I_{k})$. The conditional distribution of $\left(I_{k} - P_{R}\right)^{-1/2}U$ given $\tilde{R} = \Omega(F_{\theta}, \beta_0)^{1/2}\tilde{D}$ reads
\begin{align*}
    \left(I_{k} - P_{\tilde{R}}\right)^{-1/2}U \big| \tilde{R} \sim \mathcal{N}(0, I_{k} - P_{\tilde{R}}).
\end{align*}
We know that $\tilde{R}$ is independent of $N$ by Lemma \ref{lemma-alternative-pi-est}. Therefore, this result also holds unconditionally. Furthermore, as the rank of $I_{k} - P_{\tilde{R}}$ is $tr(I_{k} - P_{\tilde{R}}) = k - 1$, we have
\begin{align*}
    nRJ(F_n) \overset{d}{\to} \chi_{k - 1}^2.
\end{align*}
The asymptotic distribution of $nRK(F_n, \beta_0)$ and $nRJ(F_n, \beta_0)$ are independent, as $nRJ(F_n, \beta_0)$ projects on the orthogonal complement of $R(F_n)$.

Now the result follows.
\end{proof}

\newpage

\bibliographystyle{asa}
\baselineskip 17pt
\bibliography{CLRtestrob}



\end{document}